# Design and Concept Selection Analysis of Advanced Digital Stethoscope: A Systematic Approach to Medical Device Innovation

Abraham G. Taye[1,2], Sador Yonas[1], Eshetu Negash[1], Yared Minwuyelet[1], Nabiha Tofik[1]

## Abstract

In the ever-evolving landscape of medical diagnostics, this study details the systematic design process and concept selection methodology for developing an advanced digital stethoscope, demonstrating the evolution from traditional acoustic models to AI-enhanced digital solutions. The device integrates cutting-edge AI technology with traditional auscultation methods to create a more accurate, efficient, and user-friendly diagnostic tool. Through systematic product planning, customer need analysis, and rigorous specification development, we identified key opportunities to enhance conventional stethoscope functionality. The proposed system features real-time sound analysis, automated classification of heart sounds, wireless connectivity for remote consultations, and an intuitive user interface accessible via smartphone integration. The design process employed a methodical approach incorporating customer feedback, competitive benchmarking, and systematic concept generation and selection. Through a structured evaluation framework, we analyzed different distinct design concepts including traditional dual-head acoustic analogue electronic dual-head, digital single-head with condenser microphone, and Bluetooth-enabled smart stethoscope. Our selection process employed critical criteria: portability, frequency response sensitivity, transmission quality, maintenance ease, user interface simplicity, output signal quality, power efficiency, and cost-effectiveness. The final design prioritizes biocompatibility, reliability, and cost-effectiveness while addressing the growing demand for telemedicine capabilities in cardiovascular care. The project emphasizes the transition from conventional design to advanced digital solutions while maintaining a focus on practical clinical applications. Each concept was modelled using SOLIDWORKS software, enabling detailed visualization and engineering analysis. This systematic approach to concept screening and selection ensures the final design meets both current healthcare needs and future technological adaptability.

*Correspondence:
Abraham G Taye,
abraham.taye@marquette.edu
[1] Center of Biomedical Engineering, AAiT, Addis Ababa University, Addis Ababa, Ethiopia
[2] Biomedical Engineering, Marquette University, Milwaukee, WI, USA

# Table of content

# 1) Product Planning

## 1.1 Charter of project

Our motivation to design a device in the cardiovascular system is to bring an enormous effort in eradicating the fatality of cardiovascular diseases and to assist cardiologists by introducing updated medical devices.
Our group emphasis on design thinking to find the perfect point of intersection between desirability, feasibility, and viability of a cardiovascular device that can aid physicians. So as to, bring an exceptional device, we as a team took an oath to follow the following guiding principles.

- Empathize with the intended users of the product by understanding their needs and pain points
- Define the problem and generate opportunities in the human-centered way rather than a product-centered way
- Brainstorm and generate as many relevant, innovative ideas as possible
- Move forward with the best ideas by creating functional specifications
- Conduct usability tests

## 1.2 Generated opportunities:

- Remote dielectric sensing: quantifies lung fluid concentration using differences in dielectric properties of tissues.
- Bio-impedance sensors: offers valuable information about tissue or cell physiology and pathology in chronic patients.
- Cardiopulmonary exercise testing: monitors heart rate, oxygen concentration, and activity level.
- Implantable cardioverter-defibrillators with enhanced finite element computer modeling of lead.
- Arrhythmia detector and alarm: monitors an ECG and produced visible/audible signals when blood pressure falls preset.
- Blood pressure alarm
- Wireless implantable fiber-optic oximeter catheter: estimate oxygen concentration of blood.
- Create Manual/Automatic syringe actuator for an injector: timing of an injection by an angiographic or indicator injector and synchronizes the injection with the electrocardiograph signal.
- Handheld echocardiography: picks up the echoes of sound waves and transmits them as electrical impulses compact and battery operated (simplicity, portable, low-cost)
- Prosthetic vascular grafts with fewer thrombogenic surfaces.
- Mechanical valve replacement: curved leaflets to decrease pressure drops through the valve and smooth the passage of blood past the hinges that eliminate clotting issues.
- Automatic portable blood component extractor that separates blood components from centrifuged whole blood
- Cardiopulmonary bypass accessory equipment
- A ventricular bypass (assist) device
- wearable cardiac defibrillator
- Oscillometer blood volume measuring that is used to calculate the red cell mass, plasma volume, and total blood volume with 1% error.
- An external transcutaneous cardiac pacemaker (noninvasive) is used to supply a periodic electrical pulse intended to pace the heart
- A compressible limb sleeve with the flexible fabric is used to prevent pooling of blood in a limb by inflating periodically a sleeve around the limb
- A CPR Aid device with feedback provides real-time feedback to the rescuer regarding the quality of CPR being delivered to the victim

- Mechanical ventilator that automatically controls moisture level based on patient breathe moisture level (after cardiac surgery)
- Pneumatically external cardiac compressor
- Electronic stethoscope which doesn't require substantial clinical experience and good listening skills.
- Electronic stethoscope allows general practitioners to hear the sounds of auscultation clearly.
- Electronic stethoscope which allows the doctors to connect via Wi-Fi the electronic stethoscope with smartphone (and/or tablet) and while listening to the auscultation can visualize the phonogram to contrast with this double source of information its impressions
- Advanced digital stethoscope with embedded artificial intelligence for clustering and analyzing patient's condition, which allows the physician or the patient himself to analyze the sound of auscultation and health conditions.
- Electronic stethoscope which allows clinicians to apply the electronic stethoscope filter that adapts to the respiratory or cardiac sounds to collect the specific sounds. This feature facilitates and complements the patient's exploration.
- Digital stethoscope, the doctor can record and send the auscultation to another partner to receive a second opinion immediately and make a more accurate diagnosis.
- Electronic stethoscope which results in the reduction of unnecessary costs in favor of the efficiency of the health system and the quality of patient care
- High-end electronic stethoscopes allow users to adjust the frequency range they are listening for.

## 1.3 Opportunity Tournament:

Our group used a list of criteria to screen the opportunities. The criteria we used for the screening process are the following:
Then we classify the above criteria into three selection criteria in order to reduce the complexity of the screening process.

- Screening Criteria:
  - Biocompatibility
  - Raw material
  - Customer need
- Developing Promising opportunities criterions
  - Cost
  - Risk
  - Performance and reliability
- Selecting exceptional opportunities
  - Aesthetics and usability
  - Lifetime
  - Customizability
  - Market demand

Using Screening criterion, the possible 30 opportunities are going to reduce to 10 promising opportunities.

Table 1:

| **Generated opportunities** | Screening criterion A | | | Acceptance |
|---|---|---|---|---|
| | Bio compatibility | Raw material | Customer need | |
| 1. Remote dielectric sensing: | ✓ | ✗ | ✗ | Fail |
| 2. Pneumatically external cardiac compressor | ✓ | ✓ | ✗ | Fail |
| 3. An external transcutaneous cardiac pacemaker (noninvasive) | ✓ | ✓ | ✓ | Pass |
| 4. Bio-impedance sensors: | ✓ | ✓ | ✗ | Fail |
| 5. Cardiopulmonary exercise testing: | ✓ | ✗ | ✓ | Fail |
| 6. Arrhythmia detector and alarm: | ✓ | ✓ | ✗ | Fail |
| 7. Blood pressure alarm | ✓ | ✓ | ✓ | Pass |
| 8. Wireless implantable fiber-optic oximeter catheter | ✓ | ✓ | ✓ | Pass |
| 9. Automatic syringe actuator for an injector | ✓ | ✓ | ✓ | Pass |
| 10. Handheld echocardiography: | ✓ | ✗ | ✗ | Fail |
| 11. Prosthetic vascular grafts with fewer thrombogenic surfaces. | ✓ | ✓ | ✓ | Pass |
| 12. wearable cardiac defibrillator | ✓ | ✓ | ✓ | Pass |
| 13. Implantable cardioverter-defibrillators | ✗ | ✓ | ✓ | Fail |
| 14. A ventricular bypass (assist) device | ✓ | ✓ | ✗ | Fail |
| 15. Cardiovascular stent is a tube device which is intended to permanent implanted in human vessel or artificial vessel. | ✗ | ✓ | ✗ | Fail |
| 16. Heart preservation/transport system | ✗ | ✗ | ✓ | Fail |
| 17. An automatic rotating tourniquet is a device that prevents blood flow in one limb at a time. | ✓ | ✗ | ✗ | Fail |
| 18. A high-energy DC-defibrillator | ✓ | ✓ | ✗ | Fail |
| 19. A compressible limb sleeve with flexible fabric a sleeve around the limb | ✓ | ✓ | ✓ | Pass |
| 20. A CPR Aid device with feedback provides real-time feedback to the rescuer regarding the quality of CPR being delivered to the victim | ✓ | ✓ | ✗ | Fail |
| 21. Surgical vessel dilator is a device used to enlarge or calibrate a vessel. | ✗ | ✓ | ✗ | Fail |
| 22. Endovascular suturing system intended to provide fixation and sealing between an endovascular graft and the native artery | ✗ | ✓ | ✓ | Fail |
| 23. Pacemakers with piezoelectric that has better durability and portability | ✗ | ✓ | ✓ | Fail |
| 24. Cardiovascular intravascular filter for preventing pulmonary blood clots | ✗ | ✓ | ✓ | Fail |
| 25. Automatic portable blood component extractor that separates blood components from centrifuged whole blood | ✓ | ✓ | ✓ | Pass |
| 26. Oscillometer blood volume measuring device | ✓ | ✓ | ✓ | Pass |
| 27. Mechanical ventilator that automatically control moisture level based on the patients breathe moisture level ( after cardiac surgery) | ✓ | ✓ | ✓ | Pass |
| 28. Electronic stethoscope which allow the doctors to connect via Wi-Fi or Bluetooth the electronic stethoscope with smartphone | ✓ | ✓ | x | Fail |
| 29. Electronic stethoscope, the doctor can record and send the auscultation to another partner to receive a second opinion immediately and make a more accurate diagnosis. | ✓ | ✓ | ✓ | Pass |
| 30. Advanced digital stethoscope with loaded filtering and amplifying electronics and artificial intelligence | ✓ | ✓ | ✓ | Pass |

The above table shows the screening step of the opportunity tournament which is held using criterias. The choice isn't held randomly rather we use judgment of individuals and voting depends on the given criteria. These criteria help us to reduce list of opportunities into 10 promising opportunities to work with.

## 1.4 Develop promising opportunity

Using developing promising criteria possible 10 opportunities are going to reduce to 5 exceptional opportunities.

Table 2:

| **10 promised opportunities** | criterions | | | Acceptance |
|---|---|---|---|---|
| | cost | Performance & Reliability | Risk (low to medium) | |
| Blood pressure alarm | ✓ | ✓ | ✓ | Pass |
| Wireless implantable fiber-optic oximeter catheter | ✓ | | ✓ | Fail |
| Automatic syringe actuator for an injector | ✓ | ✓ | ✓ | Pass |
| An external transcutaneous cardiac pacemaker (noninvasive) | ✓ | × | ✓ | Fail |
| Mechanical ventilator that automatically control moisture level based on the patients breathe moisture level ( after cardiac surgery) | ✓ | ✓ | ✓ | Pass |
| Automatic portable blood component extractor that separates blood components from centrifuged whole blood | ✓ | ✓ | × | Fail |
| Oscillometric blood volume measuring device | × | × | ✓ | Fail |
| wearable cardiac defibrillator | ✓ | | ✓ | Pass |
| A compressible limb sleeve with flexible fabric a sleeve around the limb | ✓ | × | ✓ | Fail |
| Prosthetic vascular grafts with fewer thrombogenic surfaces. | ✓ | ✓ | × | Fail |
| Advanced digital stethoscope with loaded filtering and amplifying electronics and artificial intelligence | ✓ | ✓ | ✓ | Pass |

## 1.5 Select exceptional opportunities

Here we are selecting the last final opportunity using the following criterion, which lets the 5 promising opportunities be reduced to the last one.

Table 3:

| **Final 5 exceptional opportunities** | Screening Criteria C | | | | **Acceptance** |
|---|---|---|---|---|---|
| | Customizability | Market demand | Aesthetics and usability | Developing time | |
| Mechanical ventilator that automatically control moisture level based on the patients breathe moisture level (after cardiac surgery) | × | × | ✓ | × | Fail |
| Wireless implantable fiber-optic oximeter catheter | ✓ | ✓ | ✓ | × | Fail |
| Advanced digital stethoscope with loaded filtering and amplifying electronics and artificial intelligence | ✓ | ✓ | ✓ | ✓ | **Pass** |
| Blood pressure alarm | ✓ | × | ✓ | × | Fail |
| Automatic syringe actuator for an injector | ✓ | × | ✓ | × | Fail |

## 1.6 Evaluate and Prioritize Projects

Our chosen was developed according to horizon 2 opportunities. It is
Those are:

- ✓ Technology leadership strategy
- ✓ Cost leadership strategy
- ✓ Customer focus strategy
- ✓ Imitative strategy

## 1.7 Allocate Resource and Plan Time

Resource allocation is the identification and allocation of resources or raw materials needed for the opportunity that we have select or screen out.

Resource allocation for the advanced digital electronic stethoscope with embedded artificial intelligence are

- ✓ Lithium-Ion battery
- ✓ back up battery,
- ✓ headphones
- ✓ LED/back light
- ✓ Wireless interface system
- ✓ Battery charger
- ✓ Battery protections
- ✓ Linear voltage regulator
- ✓ Microphone
- ✓ Amplifiers
- ✓ Microcontrollers

## 1.8 Complete Pre-Project Planning

This project is approved by the 5 core team members and the mission statement is selected as
Advanced digital stethoscope

- **Timing of product introductions:** we will introduce our product at the end of two months.
- **Technology readiness**: a proven, robust technology that integrates into our products leads to produce much more quickly and reliably.
- **Market readiness:** the introduction of our product determined by our low-end product buyer and high-end product buyer.
- **Competition:** releasing of the competitive product may accelerate the development time of our product.

| Mission statement: Advanced electronics stethoscope | |
|---|---|
| Product Description | • The stethoscope operates by using Lithium-Ion battery |
| Benefit Proposition | • Allow the physician or the patient himself to analyses the sound of auscultation and health conditions- Telemedicine.<br>• High-end electronic stethoscopes allow users to adjust the frequency range they are listening for.<br>• Allows general practitioner to hear the sounds of auscultation clearly. |
| Key Business Goals | • Support advanced stethoscope in our country<br>• make the business profitable<br>• become competitive<br>• satisfy customers |
| Primary Market | • Hospitals, clinics, and NGOs<br>• Different biomedical equipment and medical device sellers<br>• Private cardiology devices importers |
| Secondary Markets | • Various biomedical instrumentations centers<br>• Private nursing homes |
| Assumptions and Constraints | • Manufactures a Powered product<br>• Expects to bring out a comfortable product<br>• Since most of the existing stethoscope in our surrounding operated manually, physicians only understood the signal |
| Stakeholders | • User<br>• Sales<br>• Distributors and resellers<br>• purchasers |

### a. Reflect on the Results and the Process:

We have identified 30 opportunities from internal and external source. We consider a final opportunity based upon the 3 selection criterions. The innovation chart is narrowly focused since we consider only the cardiovascular system. We selected numerous medical devices that can assist in the diagnosis and medication of the system. The conducted screening criterions are based on the best possible estimates of the product's success.

- ✓ As clearly stated in the above consecutive steps, the generated opportunities are 30. They all are generated from members of our group and complaints we found on the internet.
- ✓ Our innovation charter was as narrow as possible in order to erase further complications considering our planning time.

- ✓ We were considering the real opportunities depend on our passion and excitement to work on. Throughout the filtering process on the promising opportunities, we prioritized our passion and the goal we want to achieve. So, we are excited to go through the process and on the resulting opportunities.

- Is the opportunity funnel collecting an exciting and diverse set of product opportunities?
    - Yes
- Does the product plan support the competitive strategy of the firm?
    - Yes
- Does the product plan address the most important current opportunities facing the firm?
    - Yes
- Does the core team accept the challenges of the resulting mission statement?
    - Yes
- Are the elements of the mission statement consistent?
    - Yes
- Are the assumptions listed in the mission statement really necessary or is the project over-constrained?
    - Yes
- Will the development team have the freedom to develop the best possible product?
    - Yes

**Product timing table**

| | **November** | | | **December** | | | | **January** | | |
|---|---|---|---|---|---|---|---|---|---|---|
| | Week 1 | Week2 | Week3 | Week 1 | Week2 | Week3 | Week4 | Week 1 | Week2 | Week3 |
| **Product Planning** | Project duration | | | | | | | | | |
| **Identifying Customer Need** | | | Project duration | | | | | | | |
| **Product Specifications** | | | | | Project duration | | | | | |
| **Concept Generation** | | | | | | Project duration | | | | |
| **Concept Selection** | | | | | | | | | Project duration | |

How can the product planning process be improved?

- ➢ Any production plan should begin with a few questions: what products will you need to produce to meet demand? What quantity will you need to produce? And what materials will you need in order to achieve that quantity? Because sourcing for many materials requires a longer lead time than the typical customer order, most manufacturing businesses have to rely on past demand to estimate their sourcing needs in advance.

# 2)Identifying customer needs

The economic success of most firms depends on their ability to identify the needs of customers and to quickly create products that meet these needs and can be produced at low cost

## 2 .1 Gathering raw data from customers

In order to perform the duty in this step, we use the Interview method. We prepared questionnaires to ask doctors and medical staff how to look like a suitable stethoscope. We all know that there is a number of data collection methods and the most common are interviews, focus groups, and observing the product in use. So, depending on our working time and access to perform; we chosen e interview and uses it to gather the data we wanted from the customers.

The questions we prepared for the interview purpose are:

- What type of stethoscope do you need to use?
- What you do like about the existing product?
- What do you dislike about the existing product?
- What improvements do you wish to have on the product?

| Question (prompt) | Customer statements |
|---|---|
| **What type of stethoscope do you need to use?** | I need a stethoscope that has a noise filtering system |
| | I need to connect the stethoscope to my Bluetooth hearing aid |
| | I need the stethoscope to have greater strength and durability |
| | I prefer to have a display system which indicates time, date, battery level, volume and frequency. |
| | I need a stethoscope with a functional microphone that can grab analog sound waves |
| | I need a stethoscope with a sublime amplification system |
| | I want to switch back to analog from digital mode if I want to hear fidelity acoustic sounds |
| | I need a flexible tubing which i can insert into my pocket easily |
| | I prefer my stethoscope to have a volume control system. |
| | I need a stethoscope that can filter and distinguish incoming sound signals from other noises |
| | I need a stethoscope that is lightweight and handy |
| | I prefer having a stethoscope with saving, recording, and sharing capabilities |
| **What do you like about the existing product?** | Noise-canceling and volume control systems contribute greatly in a loud environment |
| | Multiple memories make it useful to share or review with peers |
| | Most people say that it is moderately lightweight and it tucks securely around the neck |
| | The electronic stethoscope is more liable to wear and tear than the acoustic stethoscope |
| | They are easy to clean and keep its sanitation |
| | It is easy to change diaphragm/bell filters without moving scopes just by pushing a button |
| **What do you dislike about the existing product?** | The battery depletes quickly if it is kept on overnight in emergency rooms(mainly ambulance operators) use it nonstop so battery life is a problem |
| | Recording time is limited to 30 seconds |
| | Sound quality is usually low ,especially for obese people and in patients with thick wall chest – a bit of weight is recommended. |
| | The price is too high and expensive |
| | Spare parts of the stethoscope are scarce |
| | The problem of tubes not working after several years |
| | Poor fits of the earbuds of the stethoscope to the human ear |
| | The battery compartment has a weak wiring connection |
| | It has some limitations to record heart sound and you must use an external device Bluetooth to |
| | transfer data to the computer |
| **What improvements you wish to have on the product?** | I want comfortable headsets that won't harm my ears after long hours of usage |
| | The stethoscope should be used easily and routinely in association with a sphygmomanometer to assess blood pressure by listening to blood flow sounds |
| | Not a single murmur of the inner body should have been missed when using the stethoscope |
| | Hygiene of stethoscope have to be a priority to avoid contaminations and accumulated dirties |
| | Stethoscopes should be attractive to eye and presented in variety colors |
| | stethoscopes features must be added to the instrument to aid professionals with mild hearing loss in either higher frequencies or lower frequencies. |
| | It would be nice if the stethoscope is small in size to be handy everywhere. |
| | I would prefer the device to work on all patients including patients with physically deformities |
| | It would be ideal for the stethoscope to tune the sound in to particular frequencies |
| | It would suit me if I can use the stethoscope in telemedicine |
| | It would be great to avoid neck burns from the stethoscope wire. |
| | I would prefer the stethoscope have a digital modality setting to be adjusted via a mobile app or share readings via Bluetooth connectivity |

## 2.2 Interpret raw data in terms of customer need.

This method involves the interpretation of the raw data gathered from the above step. We interpret customer statement into need statement by using the following guide lines:

- Express the need in terms of what the product has to do, not in terms of how it might do it
- Express the need as specifically as the raw data
- Use positive, not negative, phrasing
- Express the need as an attribute of the product
- Avoid the words must and should.

| Question (prompt) | Customer statements | Interpretation |
|---|---|---|
| **Typical use of current stethoscope** | I need a stethoscope that has a noise filtering system | The stethoscope has a filtration system on the circuitry which filters out the noise from the useful signal. |
| | I need to connect the stethoscope to my Bluetooth hearing aid | The stethoscope has access to interface with the user's mobile phone, personal computer and other electronic devices using Bluetooth. |
| | I need the stethoscope to have greater strength and durability | The stethoscope has high life expectancy with exceptional performance of its components. |
| | I prefer to have a display system which indicates time, date, battery level, volume and frequency. | The stethoscope is equipped with a display outlet of the heartbeat, date, time and battery's charge percentage. |
| | I need a stethoscope with a well functioned microphone that can grab an analog sound wave | The stethoscope has microphone transducer that changes the analog sound wave into electrical signals in terms of voltage |
| | I need a stethoscope with sublime amplification system | The stethoscope is installed with operational amplifier that has high voltage gain |
| | I want to switch back to analog from digital mode if I want to hear fidelity acoustic sounds | The stethoscope is versatile which enables users to switch from digital mode to analogue mode by just pressing ON/OFF button |
| | I need a flexible tubing which I can insert in to my pocket easily | The stethoscope has acoustic tubes made up of rubber like structure which makes its flexible |
| | I prefer my stethoscope to have a volume control system. | The stethoscope has an adjustment button for increasing and decreasing the incoming sound signals |
| | I need a stethoscope that can filter and distinguish incoming sound signals from other noises. | The stethoscope has a well embedded circuit board that can differentiate foreign artifacts and interferences using noise-cancelling technology. |
| | I prefer having a stethoscope with saving, recording, and sharing capabilities | The stethoscope has access to interface with the user's mobile phone, personal computer and other electronic devices using Bluetooth. |
| | I prefer the stethoscopes to measure vital signs of blood pressure, pulse, lung sounds, | The stethoscope delivers high-quality sound to the end user. |
| | abdominal gas, breath sounds. | |
| **Likes of current digital stethoscope** | Noise-canceling and volume control systems contribute greatly in a loud environment | The stethoscope has filtering mechanism and audio adjustment action |
| | Multiple memories make it useful to share or review with peers. | The stethoscope has options of saving, storing and recording of sound signals to ensure peer review |
| | Most people say that it is moderately lightweight and it tucks securely around the neck. | The stethoscope is lightweight and comfortable around neck. |
| | The electronic stethoscope is more liable to wear and tear than the acoustic stethoscope | The stethoscope is more liable to wear and tear. |
| | They are easy to clean and keep its sanitation | The stethoscope is easy to clean. |
| | It is easy to change diaphragm/bell filters without moving scopes just by pushing a button | The stethoscope easily switches between diaphragm and bell filters without moving scopes just by pushing a button |
| **Dislikes of current digital stethoscope** | The battery depletes quickly if it is kept on overnight - emergency rooms(mainly ambulance operators) use it nonstop so battery life is a problem | The stethoscope's battery is constructed using lithium-ion cells with have higher initial capacity of storage |
| | Recording time is limited to a restricted amount of seconds. | The stethoscope is capable of recording and rewinding for a long time. |
| | Sound quality is usually low ,especially for obese people and in patients with thick wall chest – a bit of weight is recommended. | The stethoscope has LPF which allows it to detect sounds in the low frequency zone |
| | The cost price is too expensive | The stethoscope costs fair. |
| | Spare parts of the stethoscope are scarce | Spare parts for the stethoscope are available in the market |
| | The problem of tubes not working after several years | The tubes are made of polyethylene which is flexible and durable |
| | Poor fits of the earbuds of the stethoscope to the human ear | The stethoscope is structured with adjustable headset tension. |
| | The battery compartment has a weak wiring connection | The stethoscope has a digital soldering mechanism of a wiring system. |
| | It has some limitations to record some deep sounds and connecting to computer. | The stethoscope detects a wider range of frequencies and connects to a computer easily |
| **Improvement suggestion for new products** | I want comfortable headsets that won't harm my ears after long hours of usage | The stethoscope contains optional cushion foams that relieve extended pressure from the sound |
| | The stethoscope should be used easily and routinely in association with a sphygmomanometer to assess blood pressure by listening to blood flow sounds | The stethoscope easily integrates with a sphygmomanometer cuff to listen to blood pressure |
| | Not a single murmur of the inner body should have been missed when using the stethoscope | The Stethoscope amplifies the faintest sound signals |
| | Hygiene of stethoscope have to be a priority to avoid contaminations and accumulated dirties | The stethoscope is easy to be cleanable with disinfectants, alcohol-soaked wipes, and antibacterial copper surfaces. |
| | Stethoscopes should be attractive to eye and presented in variety colors | The stethoscope comes in variety of colors which are appealing to the eye |
| | Stethoscopes features must be added to the instrument to aid professionals with mild hearing loss in either higher frequencies or lower frequencies. | The stethoscope has a robust amplification system which can multiply sound 30 times the normal sound quality |
| | It would be nice if the stethoscope is small in size to be handy everywhere. | The stethoscope allows the user to use it everywhere comfortably. |
| | I would prefer the device to work on all patients including patients with physically deformities | The stethoscope works for all patients. |
| | It would be ideal for the stethoscope to tune the sound in a specific area at a particular frequencies | Stethoscope has 3 pack modes (cardiac, lung, wide range) which allows the user to ascultate specific areas more precisely |
| | | |
| | It would be great to avoid neck burns from the stethoscope wire. | Stethoscope clips are implemented to add more comfort to the neck area |
| | I would prefer the stethoscope have a digital modality setting to be adjusted via a mobile app or share readings via Bluetooth connectivity | The stethoscope has a platform of Bluetooth connectivity for sound analysis |
| | It would suit me if I can use the stethoscope in telemedicine | The stethoscope is designed in a way that it can be further modified and integrated with other medical devices |

## 2.3 Organizing the need into hierarchy

Below we categorized hierarchical list of primary and secondary customer needs for the advanced digital stethoscope. We organized the need statement into hierarchy by using the following steps: -

- Print or write each need statement on a separate card or self-stick note.
- Elimnate redundant statements.
- Group the cards according to the similarity of the needs they express.
- For each group, choose a label.
- Consider creating supergroups consisting of two to five groups.
- Review and edit the organized needs statements.

**The redundant statements are as follows:**

- The stethoscope has a filtration system on the circuitry which filters out the noise from the useful signal.
- The stethoscope is equipped with a display outlet of the heartbeat, date, time and battery's charge percentage.
- The stethoscope is versatile which enables users to switch from digital mode to analogue mode by just pressing ON/OFF button
- The stethoscope has acoustic tubes made up of rubber like structure which makes its flexible
- The stethoscope has a well embedded circuit board that can differentiate foreign artifacts and interferences using noise-cancelling technology.
- The stethoscope has access to interface with the user's mobile phone, personal computer and other electronic devices using Bluetooth.
- The stethoscope has filtering mechanism and audio adjustment action
- The stethoscope has options of saving, storing and recording of sound signals to ensure peer review
- The stethoscope is lightweight and comfortable around neck.
- The stethoscope is more liable to wear and tear.
- The stethoscope easily switches between diaphragm and bell filters without moving scopes just by pushing a button
- The stethoscope's battery is constructed using lithium-ion cells with have higher initial capacity of storage
- The stethoscope is capable of recording and rewinding for a long time.
- The stethoscope has LPF which allows it to detect sounds in the low frequency zone
- The cost of stethoscope is fair
- The stethoscope is structured with adjustable headset tension.
- The stethoscope contains optional cushion foams that relieve extended pressure from the sound
- The instrument amplifies the faintest sound signals
- The stethoscope has a robust amplification system which can multiply sound 30 times the normal sound quality
- Advanced digital stethoscope has 3 pack modes (cardiac, lung, wide range) which allows the user to auscultate specific areas more precisely

### 2.3.1 GROUPING BY SIMILARITIES

- The Advanced digital stethoscope convenient and easy to use
  - ✓ The stethoscope is versatile which enables users to switch from digital mode to analogue mode by just pressing ON/OFF button
  - ✓ stethoscope easily switches between diaphragm and bell filters without moving scopes just by pushing a button.
  - ✓ The stethoscope is structured with adjustable headset tension.
  - ✓ The stethoscope contains optional cushion foams that relieve extended pressure from the sound

- The Advanced digital stethoscope has greater strength and durability.
  - ✓ The stethoscope is more liable to wear and tear.
- The Advanced digital stethoscope is easy to control while working with the patient
  - ✓ The stethoscope is equipped with a display outlet of the heartbeat, date, time and battery's charge percentage. .
  - ✓ Advanced digital stethoscope has 3 pack modes (cardiac, lung, wide range) which allows the user to ascultate specific areas more precisely
- The digital stethoscope allows sharing of information via Bluetooth connectivity
  - ✓ The stethoscope has access to interface with the user's mobile phone, personal computer and other electronic devices using Bluetooth.
  - ✓ The stethoscope has options of saving, storing and recording of sound signals to ensure peer review
- The digital stethoscope's battery life time is convenient for long term usage
  - ✓ The stethoscope's battery is constructed using lithium-ion cells with have higher initial capacity of storage
  - ✓ The stethoscope is capable of recording and rewinding for a long time.

- The installation of the digital stethoscope is simple and portable
  - ✓ The stethoscope is lightweight and comfortable around neck.
  - ✓ The stethoscope has acoustic tubes made up of rubber like structure which makes its flexible
- The stethoscope ha filtering system to reduce interfering sounds and noise
  - ✓ The stethoscope has a filtration system on the circuitry which filters out the noise from the useful signal.
  - ✓ The stethoscope has a well embedded circuit board that can differentiate foreign artifacts and interferences using noise-cancelling technology.
  - ✓ The stethoscope has filtering mechanism and audio adjustment action
- The digital stethoscope has signal amplification system to give high quality sound
  - ✓ The stethoscope has LPF which allows it to detect sounds in the low frequency zone
  - ✓ The stethoscope has a robust amplification system which can multiply sound 30 times the normal sound quality

- The digital stethoscope can detect a variety of frequency ranges
  - ✓ The stethoscope has a robust amplification system which can multiply sound 30 times the normal sound quality
  - ✓ The stethoscope has LPF which allows it to detect sounds in the low frequency zone

## 2.4 Establish the relative importance of the need

The hierarchical list alone does not provide any information on the relative importance that customers place on different needs. Yet the development team will have to make trade-offs and allocate resources in designing the product. A sense of the relative importance of various need is essential to making these trade-offs correctly.
We relying on the consensus of our team members based on our experience with customers, and limit the scope of customers need only about needs that are likely to give rise to difficult technical trade-offs or costly features in the product design. These are a scale of 1 to 5 how important the feature is:

1. Feature is undesirable. I would not consider a product with this feature.
2. Feature is not important, but I would not mind having it.
3. Feature would be nice to have, but is not necessary.
4. Feature is highly desirable, but I would consider a product without it.
5. Feature is critical. I would not consider a product without this feature

Indicate the feature by the number to the left if you feel that the feature is desirable, critical, undesirable, and/or not important.

__5__ The digital stethoscope is strong and durable.
___4_ The stethoscope has access to interface with the user's mobile phone, personal computer and other electronic devices.
__4__ The stethoscope has a platform of Bluetooth connectivity for sound analysis and telemedicine
__4__ The stethoscope has high life expectancy with exceptional performance of its components.
__3__ The stethoscope is equipped with a viewable outlay of heartbeat, time, date and battery values.
__4__ The stethoscope has microphone transducer that changes the analog sound wave into electrical signals in terms of voltage
__2__ The stethoscope is installed with operational amplifiers with high voltage gain.
__4__ The stethoscope is versatile which enables users to switch from digital mode to analogue mode by just pressing ON/OFF button
__4__ The stethoscope has acoustic tubes made up of rubber like structure
__5__ The stethoscope has filtering mechanism and audio adjustment action

__3__ The stethoscope has a well embedded circuit board that can differentiate foreign artifacts and interferences using noise-canceling technology.
__2__ The stethoscope is achieved more with the lightest-weight that comforts caregiver
__4__ The stethoscope has an underlying faculty to capture heart sounds and save them which can also be shared with colleagues for further assessment

## 2.5 Reflect on the result and the process

To show identified customer need process fully developed, there are some questions to ask the team include:

- ✓ Are there areas of inquiry we should pursue in follow-up interviews or surveys?
  - No
- ✓ Which of the customers we spoke to would be good participants in our ongoing development efforts?
  - End-user
- ✓ Have we interacted with all of the important types of customers in our target market?
  - Yes, we have interacted as much as we could.
- ✓ Are we able to see beyond needs related only to existing products in order to capture the latent needs of our target customers?
  - Yes
- ✓ What do we know now that we didn't know when we started? Are we surprised by any of the needs?
  - Before the project we really don't know how to manage and facilitate information gathering about customer's needs. But now we trained ourselves using those steps.
  - Ways of translation of customers need to product attribute and we translate them.
  - Ways of hierarchal organization of the need statement and organizing them.
  - Ways of rating relative importance of the need statement and we rating them.
- ✓ Did we involve everyone within our own team who needs to deeply understand customer needs?
  - Yes
- ✓ How might we improve the process in future efforts?
  - In the future we will try to involve different experts in process of improving our product.
  - By developing all guideline and criteria into good steps.

- ✓ efforts?
  - End-user
- ✓ Have we interacted with all of the important types of customers in our target market?
  - Yes, we have interacted as much as we could.
- ✓ Are we able to see beyond needs related only to existing products in order to capture the latent needs of our target customers?
  - Yes
- ✓ What do we know now that we didn't know when we started? Are we surprised by any of the needs?
  - Before the project we really don't know how to manage and facilitate information gathering about customer's needs. But now we trained ourselves using those steps.
  - Ways of translation of customers need to product attribute and we translate them.
  - Ways of hierarchal organization of the need statement and organizing them.
  - Ways of rating relative importance of the need statement and we rating them.
- ✓ Did we involve everyone within our own team who needs to deeply understand customer needs?
  - Yes
- ✓ How might we improve the process in future efforts?
  - In the future we will try to involve different experts in process of improving our product.
  - By developing all guideline and criteria into good steps.

# 3) Establishing Target specification

## 3.1 Prepare the list of metrics

Target specification need statement with relative importance

| no | need | Importance |
|---|---|---|
| 1 | The Advanced digital stethoscope has greater strength and durability. | 5 |
| 2 | The stethoscope has high life expectancy with gigantic performance of its components | 5 |
| 3 | The stethoscope is more liable to wear and tear | 5 |
| 4 | The stethoscope power is convenient | 5 |
| 5 | The battery does not deplete quickly if it kept on overnight emergency room | 5 |
| 6 | Advanced digital stethoscope has 3 pack modes (cardiac, lung, wide range) which allows the user to auscultate specific areas more precisely | 4 |
| 7 | The battery is rechargeable | 4 |
| 8 | Advanced digital stethoscope has multiple memories which makes it useful to share or review with peers. | 4 |
| 9 | The stethoscope is capable of recording and rewinding for a long time. | 4 |
| 10 | The stethoscope allows the user to use it everywhere comfortably | 4 |
| 11 | The device works on all patients including physically deformed patients | 4 |
| 12 | The Advanced digital stethoscope easy to set up and use | 3 |
| 13 | The Advanced digital stethoscope is easy to clean with alcohol. | 3 |
| 14 | The display system which indicates time date, battery level, volume and frequency is really handy. | 3 |
| 15 | stethoscope is easy to change diaphragm/bell filters without moving scopes just by pushing a button. | 3 |
| 16 | Advanced digital stethoscope is lightweight and it tucks securely around the neck | 3 |
| 17 | stethoscope have a digital modality connectivity | 3 |
| 18 | The Advanced digital stethoscope is easy to see the volume | 2 |
| 19 | The installation of the digital stethoscope is simple and portable | 2 |
| 20 | The Advanced digital stethoscope is easy to control while working with the to the patient | 2 |

We prepare our metrics list using the following guidelines:

- Metric should be complete
- Metric should be dependent variables
- Metric should be practical
- Some needs cannot easily be translated into quantifiable metrics
- The metrics should include the popular criteria for comparison in the market place

Table 1: list of metrics

| Metricno | Metrics | Imp. | Units |
|---|---|---|---|
| 1. | Time of assembly | 1 | minute |
| 2. | Sound volume adjustment | 3 | Decibel |
| 3. | Connectivity with other external devices | 3 | list |
| 4. | Turning of the diaphragm to bell filter. | 3 | subj |
| 5. | The resolution of the display system | 2 | Dots/inch |
| 6. | Power consumption. | 3 | Watts |
| 7. | Strength and durability | 5 | $N/m^2$ |
| 8. | Life span expectancy. | 4 | years |
| 9. | Sound amplification and filtration range | 5 | Hz |
| 10. | Length of the sound tubing. | 3 | Cm |
| 11. | Total mass of the digital stethoscope | 3 | g |
| 12. | cost for maintenance | 2 | USD |
| 13. | Flexibility of wiring system. | 3 | m/N |
| 14. | Sound pressure level of ear piece | 4 | dB |
| 15. | Maximum capacity of memory used. | 2 | Gb |
| 16. | Running time of the stethoscope's battery | 4 | Hr |
| 17. | cost for manufacturing. | 3 | USD |
| 18. | The stethoscope makes user independent. | 3 | Subj |
| 19. | Sensitivity of stethoscope | 4 | Radian/meter |
| 20. | Special spare parts for maintenance | 2 | list |
| 21. | Operating temperature of stethoscope | 3 | K |
| 22. | Bell mode and diaphragm amplification capacity | 4 | hz |
| 23. | Patient heart rate measurement range | 4 | Beat /min |
| 24. | Automated power management system (on/ off) | 2 | subj |
| 25. | Bluetooth modality distance range | 3 | m |
| 26. | Relative humidity | 2 | % |

## 3.2 Need Metrics Matrix

| | 1) Time of assembly | 2) Sound volume adjustment | 3) Connectivity with other devices | 4) Turning of the diaphragm to bell filter | 5) Resolution of the display system | 6) Power consumption | 7) Strength and durability | 8) Life span expectancy | 9) Sound amplification and filtration range | 10) Length of sound tubing | 11) Total mass of the digital stethoscope | 12) Cost for maintenance | 13) Flexibility of wiring system. | 14) Sound pressure level of ear piece | 15) Maximum capacity of memory used | 16) Running time of the stethoscope battery | 17) Cost of manufacturing | 18) The stethoscope makes user independent | 19) Sensitivity of stethoscope | 20) Special spare part for maintenance | 21) Operating temperature of stethoscope | 22) Bell mode and diaphragm amplification | 23) Patient heart rate measurement range | 24) Automated power management system(on/off) | 25) Bluetooth modality distance range | 26) Relative humidity of stethoscope |
|---|---|---|---|---|---|---|---|---|---|---|---|---|---|---|---|---|---|---|---|---|---|---|---|---|---|---|
| The Advanced digital stethoscope easy to set up and use | | | * | | | | | | | | * | | | | | | | | | * | * | | | | | * |
| The Advanced digital stethoscope is easy to clean with alcohol. | | | | | | | * | | | | | | | | | | | * | | | | | | | | * |
| The Advanced digital stethoscope is easy to see the volume | | * | | | | | | | | | | | | | | | | * | | | | | | | | |
| stethoscope is easy to change diaphragm/bell filters without moving scopes just by pushing a button. | | | | * | | | | | | | | | | | | | | * | | | * | | | | | |
| 5. The display system which indicates time date, battery level , volume and frequency is | | | | | * | | | | | | | | | | | | | | | | | | | | | |
| really handy. | | | | | | | | | | | | | | | | | | | | | | | | | | |
| The Advanced digital stethoscope has greater strength and durability. | | | | | | | * | * | * | | | | * | | | | | * | | | | | | | | |
| The stethoscope has high life expectancy with gigantic performance of its components. | | | | | | | * | * | * | | | | * | | | | | | | * | | | | | | |
| The stethoscope is more liable to wear and tear. | | | | | | | * | * | | | | * | | | | | | * | | | | | | | | |
| The Advanced digital stethoscope is easy to control while working with the to the patient | * | | | | | | | | | | | | | | | | | * | | | | | | | | |
| The display system which indicates time date, battery level , volume and frequency is really handy | | | | | * | | | | | | | | | | | | | * | | | | | | * | | |
| Advanced digital stethoscope has multiple memories which makes it useful to share or review with peers. | | | | | | | | | | | | | | | * | * | | | | | | | | * | | |
| Advanced digital stethoscope has 3 pack modes (cardiac, lung, wide range) which allows the user to ascultate specific areas more precisely | | | | * | | | | | * | | | | | * | | | | | | | | | * | | | |
| The device works on all patients | * | | | | | | | | | | | | | | | | | * | | | | | | | | |
| including physically deformed patients. | | | | | | | | | | | | | | | | | | | | | | | | | | |
| stethoscope have a digital modality connectivity | * | | * | | | | | | | | | | | | | | | | | * | | | | | * | |
| The Advanced digital stethoscope power is convenient | | | | | | * | * | * | | | | | | | | * | | | * | | | | | * | | |
| The Advanced digital stethoscope battery do not turn off when i touched the patient | | | | | | | | | | | | | | | | | | | * | | | | | | | |
| 17. The battery is rechargeable | | | | | | * | * | * | | | | | | | | * | | | | | | | | | | |
| The battery do not deplete quickly if it is kept on overnight-emergency rooms | | | | | | | * | * | | | | | | | | * | | | | | | | | | | |
| The stethoscope is capable of recording and rewinding for a long time. | | | | | | | | | | | | | | | * | | | | | | | | | * | | |
| The installation of the digital stethoscope is simple and portable | * | | | | | | | | | | | | * | | | | | | | | | | | | | |
| Advanced digital stethoscope is lightweight and it tucks securely around the neck | * | | | | | | | | | * | * | | * | | | | | * | | | | | | | | |
| The stethoscope allows the user to use it everywhere comfortably. | | | * | | | | | | | | | | | | | | | * | | | | | | | | |

### 3.3 Competitive benchmarking information

Table 3: competitive benchmarking chart based on metrics

| No. | Metrics | Imp. | units | Acoustic stethoscope | Digital stethoscope | 3M littmann digital stethoscope |
|---|---|---|---|---|---|---|
| 1. | Time of assembly | 1 | minutes | none | Several minutes | none |
| 2. | Sound volume adjustment | 3 | Decibel | 80 | 70-80 | 70-75 |
| 3. | Connectivity with other external devices | 3 | list | none | available | available |
| 4. | Turning of the diaphragm to bell filter. | 3 | Degree | none | 180 | 180 |
| 5. | The resolution of the display system | 2 | Dots/inch | none | 100 | none |
| 6. | Power consumption. | 3 | Watts | none | 4 watt | 6 watt |
| 7. | Strength and durability | 5 | $N/m^2$ | weaker | Strong | Strong |
| 8. | Life span expectancy. | 4 | years | Max 2 yr | 5 yrs | 3 yrs |
| 9. | Sound amplification and filtration range | 5 | Hz | 689- 2584 Hz | 40x | 24x |
| 10. | Length of the sound tubing. | 3 | inch | 22-31 | ~27 | 22-27 |
| 11. | Total mass of the digital stethoscope | 3 | gram | ~180 | 165-185 | 175 |
| 12. | cost for maintenance | 2 | USD | none | | none |
| 13. | Flexibility of wiring system. | 3 | m/N | flexible | Much flexible | stiff |
| 14. | Sound pressure level of ear piece | 4 | dB | none | 80-1 | |
| 15. | Maximum capacity of memory used. | 2 | Gb | none | 2 | none |
| 16. | Running time of the stethoscope's battery | 4 | Hrs | none | 48 | 60 |
| 17. | cost for manufacturing. | 3 | USD | 150-190 | 315 | ~354 |
| 18. | The stethoscope makes user independent. | 3 | Subj | none | yes | yes |
| 19. | Sensitivity of stethoscope | 4 | Radian/meter | ~45.7- 60.6% | 93-94.4% | 56- 80% |
| 20. | Special spare parts for maintenance | 2 | list | none | available | none |
| 21. | Operating temperature of stethoscope | 3 | Fahrenheit | -32- 122 | -25 -110 | -22 - 104 |
| 22. | Bell mode and diaphragm amplification capacity | 4 | Hz | none | 10-5000 | 20-2000 |
| 23. | Patient heart rate measurement range | 4 | Beat/min | 50-100 | 40- 175 | 30-199 |
| 24. | Automated power management system(on/ off) | 2 | subj | none | available | available |
| 25. | Bluetooth modality distance range | 3 | m | none | 12 | 10 |
| 26 | Relative humidity | 2 | % | none | 15-93 | 15- 93 |

Table 4: competitive benchmarking chart based on perceived satisfaction of customers

| No. | Metrics | Imp. | Acoustic stethoscope | Digital stethoscope | 3 M Littmann stethoscope |
|---|---|---|---|---|---|
| 1 | Time of assembly | 1 | ● | ●●● | ●●●● |
| 2 | Sound volume adjustment | 3 | - | ●●● | ●●● |
| 3 | Connectivity with other external devices | 3 | - | ●●●● | ●●● |
| 4 | Turning of the diaphragm to bell filter. | 3 | - | ●●● | ●●● |
| 5 | The resolution of the display system | 2 | - | ●●● | ● |
| 6 | Power consumption. | 3 | - | ●●●● | ●●● |
| 7 | Strength and durability | 5 | ●●● | ●●● | ●●● |
| 8 | Life span expectancy. | 4 | ●● | ●●●● | ●●● |
| 9 | Sound amplification and filtration range | 5 | ● | ●●●● | ●●● |
| 10 | Length of the sound tubing. | 3 | ●●● | ●●● | ●●● |
| 11 | Total mass of the digital stethoscope | 3 | ●● | ●●●● | ●●● |
| 12 | cost for maintenance | 2 | - | ●●● | - |
| 13 | Flexibility of wiring system. | 3 | ●●● | ●●●● | ● |
| 14 | Sound pressure level of ear piece | 4 | ● | ●●● | ●● |
| 15 | Maximum capacity of memory used. | 2 | - | ●●●● | - |
| 16 | Running time of the stethoscope's battery | 4 | - | ●●● | ●●●● |
| 17 | cost for manufacturing. | 3 | ●●● | ●● | ●● |
| 18 | The stethoscope makes user independent. | 3 | ● | ●●●● | ●● |
| 19 | Sensitivity of stethoscope | 4 | ●● | ●●● | ●●● |
| 20 | Special spare parts for maintenance | 2 | - | ●●●● | - |
| 21 | Operating temperature of stethoscope | 3 | ● | ●●●●● | ●●●●● |
| 22 | Bell mode and diaphragm amplification capacity | 4 | - | ●●●●● | ●●● |
| 23 | Patient heart rate measurement range | 4 | ● | ●●● | ●● |
| 24 | Automated power management system(on/ off) | 2 | - | ●●●● | ●●●● |
| 25 | Bluetooth modality distance range | 3 | - | ●● | ●● |
| 26 | Relative humidity | 2 | ●●● | ●● | ●● |
| | | | | | |

## 3.4 Set Ideal and marginal values

We are setting our ideal and marginal accepted value by using five mechanisms:

- At least X
- At most X
- Between X and Y
- Exactly X
- A set of discrete values

| No. | Metrics | Imp. | units | Marginal values | Ideal values |
|---|---|---|---|---|---|
| 1 | Time of assembly | 1 | minutes | 10 -15 minutes | Less than 10 min |
| 2 | Sound volume adjustment | 3 | Decibel | 75 -80 db | 80 db |
| 3 | Connectivity with other external devices | 3 | list | available | availa ble |
| 4 | Turning of the diaphragm to bell filter. | 3 | Degree | 160 to 180 degree | 180 degree |
| 5 | The resolution of the display system | 2 | Dots/inch | 90 to 100 | 100 |
| 6 | Power consumption. | 3 | Watts | Bellow 4 watts | 4 watts |
| 7 | Strength and durability | 5 | $N/m^2$ | Good strength | Excellent |
| 8 | Life span expectancy. | 4 | years | Bellow 5 years | >= 5 years |
| 9 | Sound amplification and filtration range | 5 | Hz | 30-35x | 40x |
| 10 | Length of the sound tubing. | 3 | inch | 20 to 25 inch | 27 inch |
| 11 | Total mass of the digital stethoscope | 3 | gram | 150 to 170 g | 175 |
| 12 | cost for maintenance | 2 | USD | $10- $100 | $100 |
| 13 | Flexibility of wiring system. | 3 | m/N | Much flexible | Ideally flexible |
| 14 | Sound pressure level of ear piece | 4 | dB | 50 to 70 db | 80db |
| 15 | Maximum capacity of memory used. | 2 | Gb | 2gb | 4gb |
| 16 | Running time of the stethoscope's battery | 4 | Hrs | 48 hours | 60 hours |
| 17 | cost for manufacturing. | 3 | USD | $300 | $315 |
| 18 | The stethoscope makes user independent. | 3 | Subj | Yes | Yes |
| 19 | Sensitivity of stethoscope | 4 | Rad/meter | 93-94.4% | 98% |
| 20 | Special spare parts for maintenance | 2 | list | available | available |
| 21 | Operating temperature of stethoscope | 3 | F | 0-50 degree | -21 -100 degree |
| 22 | Bell mode and diaphragm amplification capacity | 4 | hz | 20-4000 | 10-50000 |
| 23 | Patient heart rate measurement range | 4 | Beat/min | 40-175 bpm | 20- 200bpm |
| 24 | Automated power management system(on/ off) | 2 | subj | available | available |
| 25 | Bluetooth modality distance range | 3 | m | 12 meter | 15 meate r |
| 26 | Relative humidity | 2 | % | 15 -93 | 0-100 |

Table 5: the target specifications

### 3.5 Reflect on the result and the process

Our team take some iteration to agree on our target's specification. Reflection after each iteration helps to ensure that the results areconsistent with the goals of our project and Questions to consider include:

- Are members of the team "gaming"? Yes
- Should the team consider offering multiple products or at least multiple options for theproduct in order to best match the particular needs of more than one market segment,
- or will one "average" product suffice? Yes
- Are any specifications missing? No
- Do the specifications reflect the characteristics that will dictate commercial success? Yes

# 4)Concept Generation

## 4.1 System Overview

The Advanced digital Stethoscope picks up sounds, such as heart and lung sounds, from a patient's body. After amplification and filtering, the sounds are sent to the physician through a binaural headset for proper diagnosis. The viewable hardware components are listed and indicated as follows:

- Bell and Diaphragm mode
- Microphones
- Controlling buttons (volume, power…)
- Battery
- Wireless system
- Tubing system
- Output visualization system (LCD screens)
- Earpieces

The system must be able to

- Use its microphones to capture the heart and lung sounds which occur at frequencies between 20 and 150 Hz and between 50 and 2500 Hz respectively, as well as filter out the ambient noise to reduce it to a maximum of 10% of the total measured incoming signal
- Take the captured audio input and determine if it corresponds to an abnormal heart or lung condition with an accuracy of ~90%
- Detect an abnormality in the heart or lung sounds within four seconds once the incoming signals have been filtered

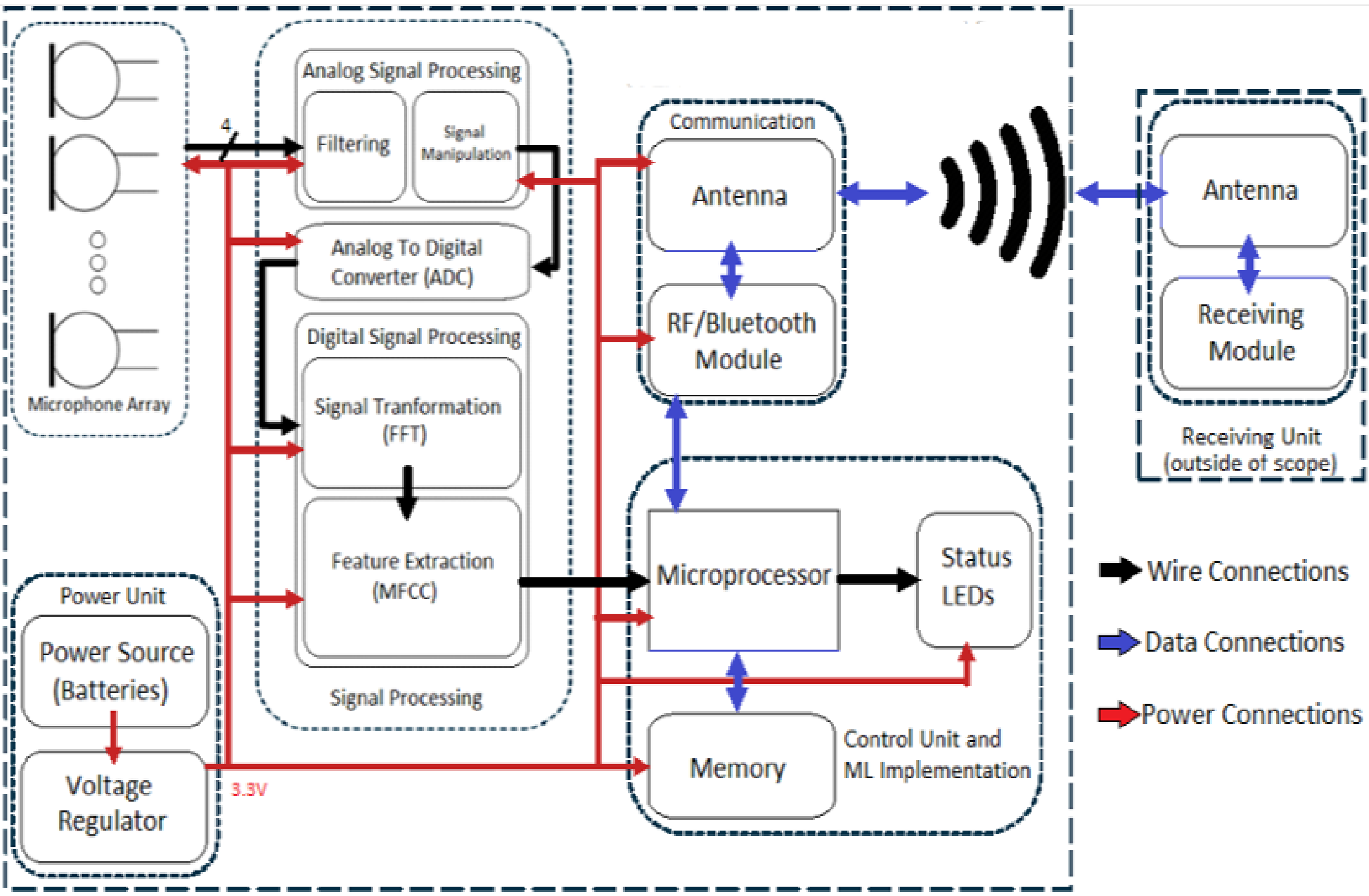

*Fig 1: Overall block diagram of audio- signal path.*

The block diagram design described in above figure fulfills the high-level requirements described and follows the path of the audio signals. First, the microphone array and associated analog and digital signal processing fulfills the system's ability to sense the sounds produced by the heart and lungs. Second, the microprocessor and memory (parts of the control unit) will allow the implementation of the machine learning algorithms required to detect conditions of the organs. Finally, the wireless capability will enable the system to warn the user (or doctor) of any detected issues.

### 4.1.1 Design Criterions and considerations

#### Signal processing phases

- **Analog signal processing:**
  Electronic stethoscopes require the conversion of acoustic sound waves obtained through the chest piece into electronic signals which are then transmitted through uniquely designed circuitry and processed to filtration and amplification for optimal listening. The analog signal processing takes care of filtering out noise coming from outside the frequency range for which we are interested in, between 20 and 150 Hz for the heart and between 50 and 2500 Hz for the lungs. This sub-unit takes its input from the microphones and manipulates the signals before sending them to the Analog-To-Digital Converter (ADC).

- **Digital signal processing**:
  The ADC will then take the output from the analog signal processing unit and convert it into digital readings. These will allow for the processing required that cannot be done through analog signals alone. The conversion to digital enables the transformations for features to be generated (MFCC). The analog to digital converter will have to be quick and have a large enough accuracy to prevent bias. It will also require a good bit-width.

- **AI Signal processing**:
  - The digital signal processing unit handles the processing of the converted and filtered signals into meaningful data to be used by the control unit (microcontroller) to detect and diagnose abnormalities. This unit will have to do this quickly and reliably. It will either be done through a dedicated microprocessor that will handle the computations.
  - The microprocessor will be the unit which does all the computations necessary for the filtering the incoming signals, converting to Mel Frequency Cepstrum Coefficients (MFCC), and implementing a machine learning algorithm to do the diagnosing. It will also handle the preparation of the data for sending through the communication unit and will choose the data to be stored in the memory.
  - The overall idea of this unit is to determine if a person has an abnormality in their heart or lungs using the k-nearest neighbor (k-NN) algorithm. While there are many other machine learning algorithms to choose from, the k-NN is easy to implement, its accuracy is relatively high, and only k and the distance metric need to be defined.

**Battery Management**

The power unit provides the power required to operate all the other components in the system. The system is comprised of components which operate at 3-5V. As such, it should be possible to use a single power line to power all the components of the system.
The battery we used is a single-cell Li+ battery. It's rechargeable and with high lifespan. A battery charger is required with a normal insertable USB-B cable. For the reason of the battery is removable, then authentication is also required for safety and aftermarket management

## MICROPHONES

The microphone array acts as the system's sensors. The sensors are the interface between the organs we are monitoring and the system we are designing.

In order to determine the best microphone to use in our final design, three different types were analyzed and compared. The three types of microphones that were considered for use were a condenser microphone, a fiber optic microphone, and a Micro Electrical-Mechanical System (MEMS) microphone.

- The first option, a condenser microphone, works by means of a capacitor, which converts acoustical energy into electrical energy
- The second microphone option was a fiber optic microphone. Fiber optic microphones work by sensing changes in light intensity, rather than changes in capacitance or magnetic fields like traditional microphones. Light from a laser source travels through an optical fiber, where it illuminates the surface of a reflective diaphragm at the tip of the microphone.
- The final microphone considered was a MEMS microphone, is the name given to very small mechanical devices driven by electricity. The microphone element consists of an impedance converter and an output amplifier transmitting a digital audio output signal recorded from the microphone head.

## Memory

The memory will be both embedded and external. Due to the sensitivity of the type of data, medical data, it is crucial to store the history of the patient. Therefore, there needs to be a means of permanently storing the patient history gathered by the device.

### Display and Backlighting

The advanced electronic stethoscope contains buttons and LED indicators. Backlighting for the display is required because the ambient lighting during the procedure is often at a low level. The small display requires just one white light-emitting diode (WLEDs) controlled by an LED driver.

### Data Storage and Transfer system

Once the captured sound is converted to an analog voltage, it can be sent out through communication unit interfaces either *Wired* or *Wireless* modalities.

## Wireless

For audio transfer, three wireless signals were considered: Bluetooth chipsets, an FM transmitter, and Pure Path wireless.

- Bluetooth chipsets utilize low-power radio frequency transmission. They are extremely small and relatively inexpensive. However, they are more difficult to integrate into a circuit, as they are inflexible in their implementation. A circuit must be built around the Bluetooth chip, increasing its complexity.
- The second wireless option analyzed was an FM transmitter. This method would be similar to sending the audio from an iPod to the radio of a car. An FM transmitter system would be easy to implement, but would use significantly more power. The transmitter is larger than desired, and the high power necessary to run it would require additional bulky batteries attached to the stethoscope. The problem of ensuring the FM signal was not in use would also have to be considered.
- The final wireless option was TI's Pure Path Wireless system. Pure Path Wireless sends an uncompressed digital audio signal over a strong radio frequency link. Pure Path was designed solely for audio transmissions, so the sound quality is excellent. This system is available in development kits, making it easy to implement. The kits also come with rechargeable batteries with a 22-hour battery life.

## 4.2 Clarifying the problem

### 1. Understanding the problems

The increasing mortality rate due to cardiovascular diseases, the need for stethoscope in telemedicine, the feasibility of cardiac and accurate bowel sounds auscultation, and the technology update to conventional acoustic stethoscope give rise to the preventive and diagnostic measures with the aid of our Advanced Digital Stethoscope.

**From mission statement**

- The advanced digital stethoscope will use rechargeable Lithium-Ion Battery
- The advanced digital stethoscope will have an LCD display system that shows frequency range, sound volume, and battery level
- The advanced digital stethoscope will use a Bluetooth connection to transmit sound data for record and save auscultation sessions
- The advanced digital stethoscope will use audio headphones to listen to heart and lung sounds
- The advanced digital stethoscope will produce high-quality sound signal output with the aid of machine learning
- The advanced digital stethoscope will have a small size of diaphragm and bell that can fit in the palm hand
- The advanced digital stethoscope will be applied in telemedicine
- The advanced digital stethoscope will be durable and long-lasting
- The advanced digital stethoscope will be equipped with active noise cancellation to reduce interferences
- The advanced digital stethoscope will highly amplify sound signals

**From customer need**

- The advanced digital stethoscope is versatile that can be used in adult patients and pediatrics patients.
- The advanced digital stethoscope uses tunable chest pieces for low and high frequencies sounds
- The advanced digital stethoscope will be easy to clean to avoid the risk of COVID-19 infections.
- The advanced digital stethoscope is easy to use and comfortable
- The advanced digital stethoscope is light-weight and handy
- The advanced digital stethoscope is cost-effective
- The advanced digital stethoscopes have connections to hearing aid streamers for clinicians with hearing loss problems.
- The advanced digital stethoscope has a pocket carrying case
- The advanced digital stethoscope will have an LCD display system that shows frequency range, sound volume, and battery level
- The advanced digital stethoscope will toggle between analog and digital listening modes
- The advanced digital stethoscope will be equipped with volume control for easy increase and decrease of sound levels
- The advanced digital stethoscope will have soft-sealing ear tips for acoustic seal and comfortable fit

**From target specifications**

- Running time of the battery exceeds 48 hours
- Bluetooth length range varies from 10 to 12 meters
- Sound signals amplified up to 40 times of the original incoming sound

- Total mass is less than 200 grams
- Five years of life expectancy
- Monitor resolution display of 90 to 100 dots/inch
- Length of tubing ranges from 25 to 30 inches
- Operating temperature of $-25^0$ up to $110^0$F
- Storage capacity up to 2GB
- Measurement of patient heart rate ranging from 40 up to 175 beats per min

### 4.2.1 Decompose a Complex Problem into Simpler Subproblems

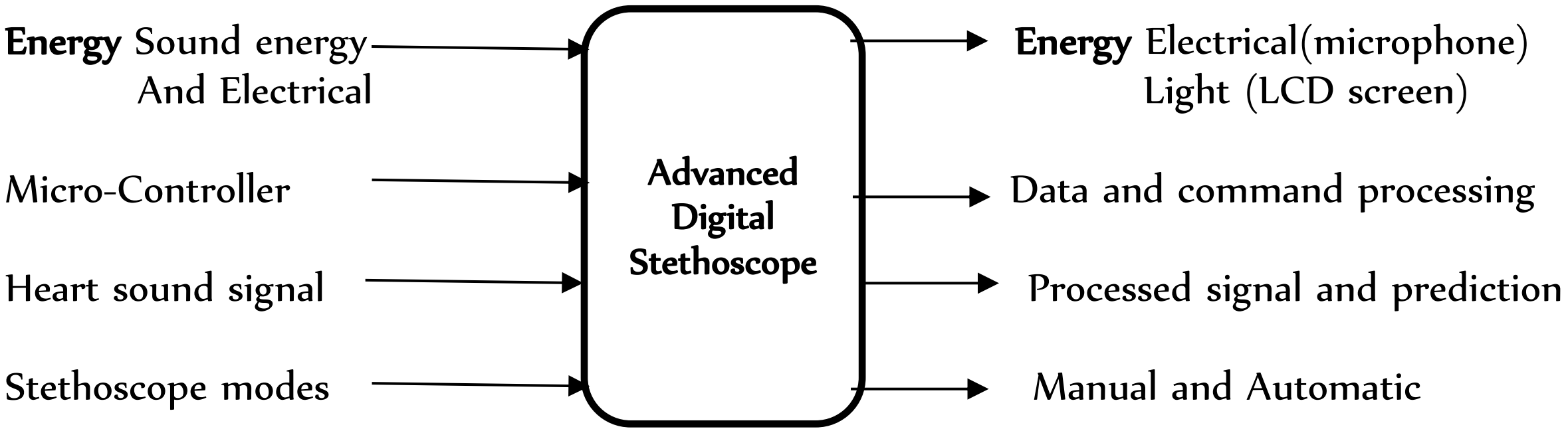

*Fig 2.1. Overall black box of subproblems*

The next step in functional decomposition is to divide the single black box into subfunctions to create a specific description of elements of the product might do in order to implement the overall function of the product

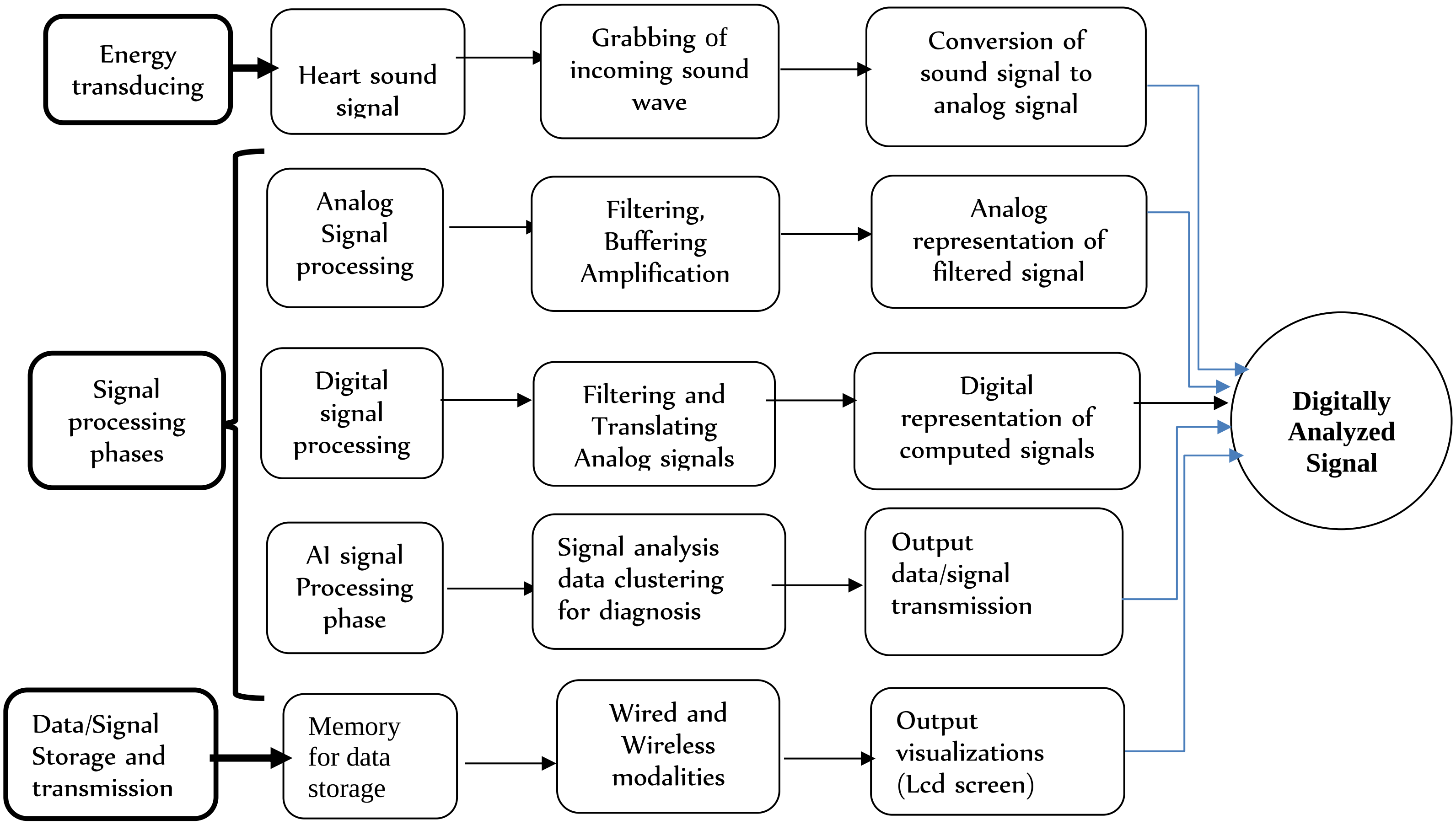

*Fig 2.2 decomposition digaram*

### 4.2.2 Focus Initial Efforts on the Critical Subproblems

The critical problems identified by the teams are

- Frequency response of microphones (sensitivity)
- Signal processing quality (phases)
- Stethoscopes head modes (dual or single)
- Improper application of telemedicine (communication interfaces)
- Output signal visualization

## 4.3 Search Externally

The external search for solutions is essentially an information-gathering process aimed at finding existing solutions to both the overall problem and the sub-problems identified during the problem clarification step finding existing solutions from:

- **Interview Lead users**
    - The frequency response of the microphone should allow for the user to hear the heart and lung sounds as much as possible.
    - Telemedicine: Proper orientation of the patient on the technical skills of the stethoscope shall be used.
- **Benchmarking**
    - Competitive products: The new 3M Littmann CORE Digital Stethoscope offers clinicians access to both analog and digital auscultation options and connects to Eko's software and AI algorithms1 to help clinicians better interpret sounds and detect heart murmurs
- **Technical lecture and journals**
    - Cardiology single head stethoscopes have a pressure-sensitive tuneable head that functions as both a diaphragm and bell depending on the applied pressure. This tuneable diaphragm allows for easy shift between high and low frequency sounds.
- **Experts**
    - While the dual head stethoscope is commonly used in healthcare, the single head stethoscope is still widely preferred among certain practitioner due it's tuneable diaphragm and ease of use.

## 4.4 Search Internally

Internal search is the use of personal and team knowledge and creativity to generate solution concepts. Often called **brainstorming**, this type of search is internal in that all of the ideas to emerge from this step are created from knowledge already in the possession of the team.

Here are some of the solutions our team generated for the subproblems of critical problems, these are

(1) Microphone's sensitivity problem,
(2) Signal processing quality related problems,
(3) Stethoscope chest piece alignment problems,
(4) Improper application of telemedicine and signal transmission related problems and
(5) Signal output visualization related problems.

1) **A solution to sensitivity problems of stethoscope:**
    - From our group's perspective we try to categorize the type of transducers and microphones based on their frequency responses
        - Piezoelectric transducers
        - Capacitive transducers
        - Optical transducers
        - Condenser microphone
        - Fiber optics microphone
        - MEMS microphone

2) **A solution to signal processing qualities.**
    - From our group's perspective we try to categorize the phase of signal processing process based on their electronics component involved.
        - Amplification and noise filtration;
        - Analog to digital conversion for digital representation of signals:
        - Machine learning enabled signal processing for diagnosis the abnormalities:

3) **A solution to stethoscope head mode system**
    - We try to categorize the type of output signal visualization based on embedded interfaces
        - A dual head model: includes the diaphragm (for high frequencies) and the bell (for low frequencies
        - A single head model: includes the diaphragm and the bell on one side, and which one you hear just depends on the pressure

4) **A solution to signal transmission problems and Improper application of telemedicine**
    - We try to categorize the type of signal communication interfaces based on their transmission qualities.
        - Collaborative work of using telehealth stethoscope between the doctor and patients' nurse
        - Usage of modern wireless communication technologies.
        - Bluetooth interface (wireless) to share with other physicians
        - FM transmitter (wireless) for better telemedicine
        - Pure-path system(wireless)
        - Tubing system for real time signal transmission
        - Fiber Optics wiring system for faster digital signal transmission

5) **A solution to output visualization system**
    - We try to categorize the type of output signal visualization based on embedded interfaces
        - Embedded high quality LCD system
        - Embedded LCD system and wireless devices for further analysis
        - Embedded seven segment LED for numbered data.

## 4.5 Explore Systematically

Systematic exploration is aimed at navigating the space of possibilities by organizing and synthesizing these solution fragments.

It's dependent on the result of the external and internal search activities that the team collected of concept fragments solutions to the subproblems.

Our team focused on the energy transducer types, bell and diaphragm alignment (chest piece modes), stethoscope signal modes, signal processing level, output signal transmission and visualization as follow:

### 4.5.1 Concept classification trees

- The concept classification tree is used to divide the entire space of possible solutions into several distinct classes that will facilitate comparison and pruning.

- Here are some concept classification trees:

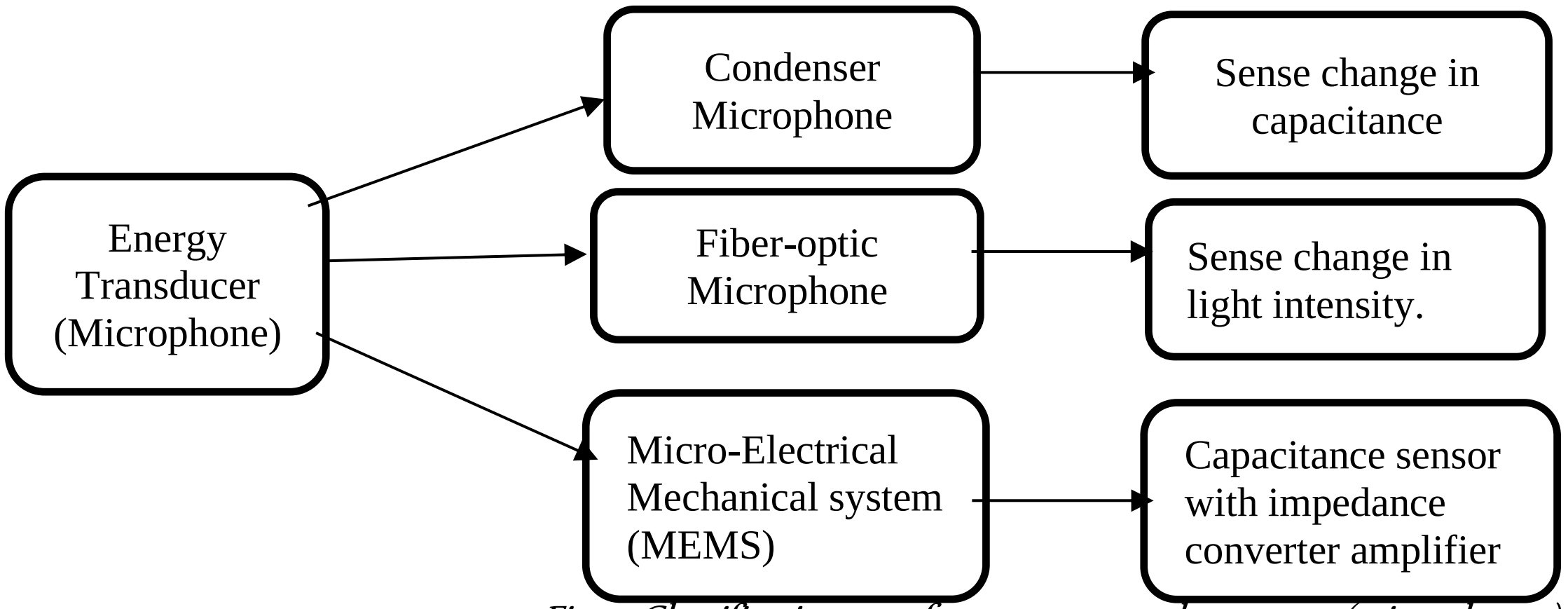

*Fig 3. Classification tree for energy transducer types(microphones)*

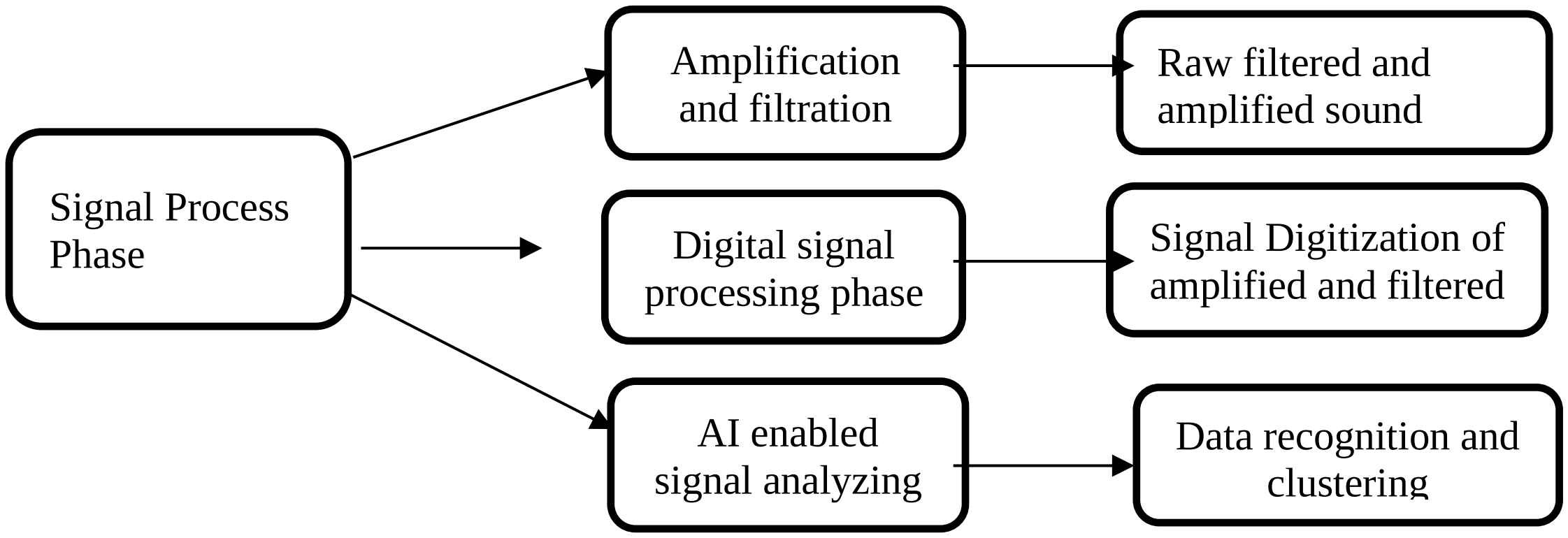

*Fig 4. Classification tree for signal processing unit capacities*

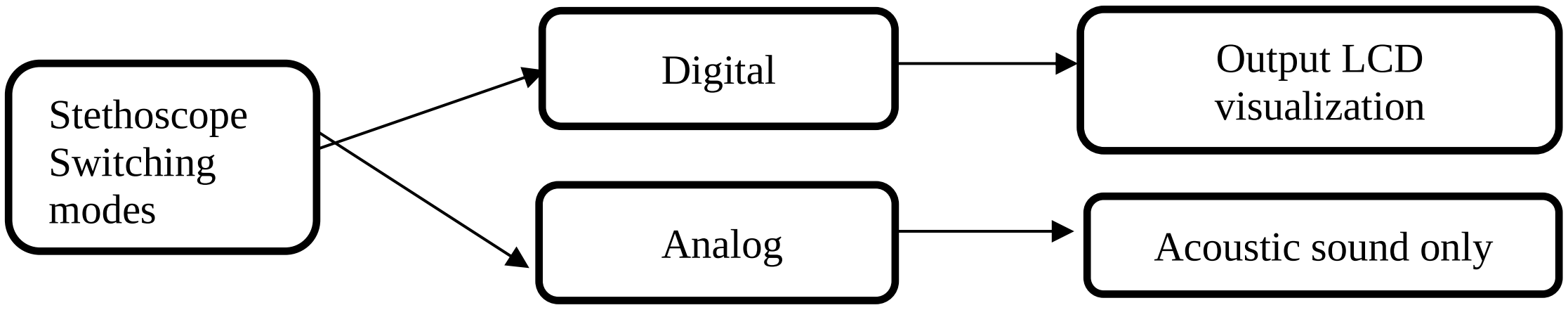

*Fig 5. Classification tree for stethoscope switching modes*

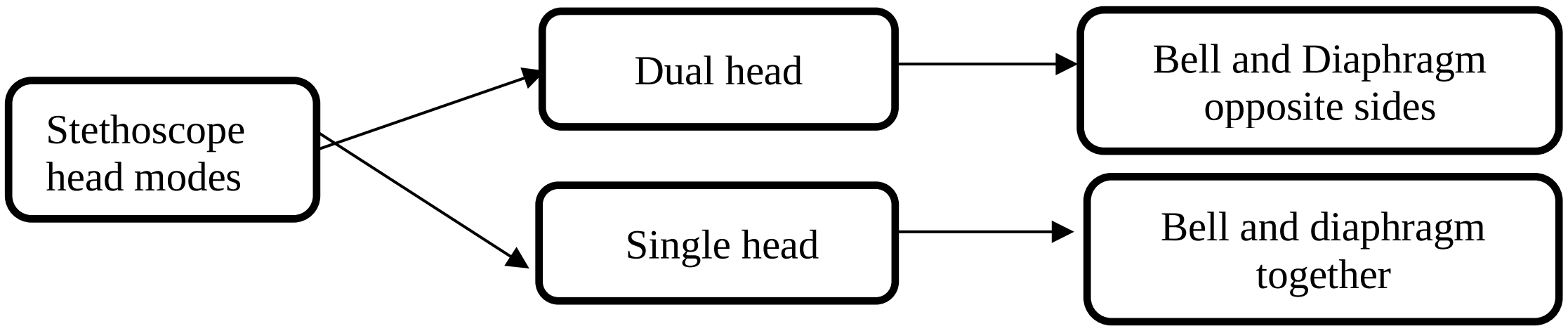

*Fig 6. Classification tree for stethoscope head modes.*

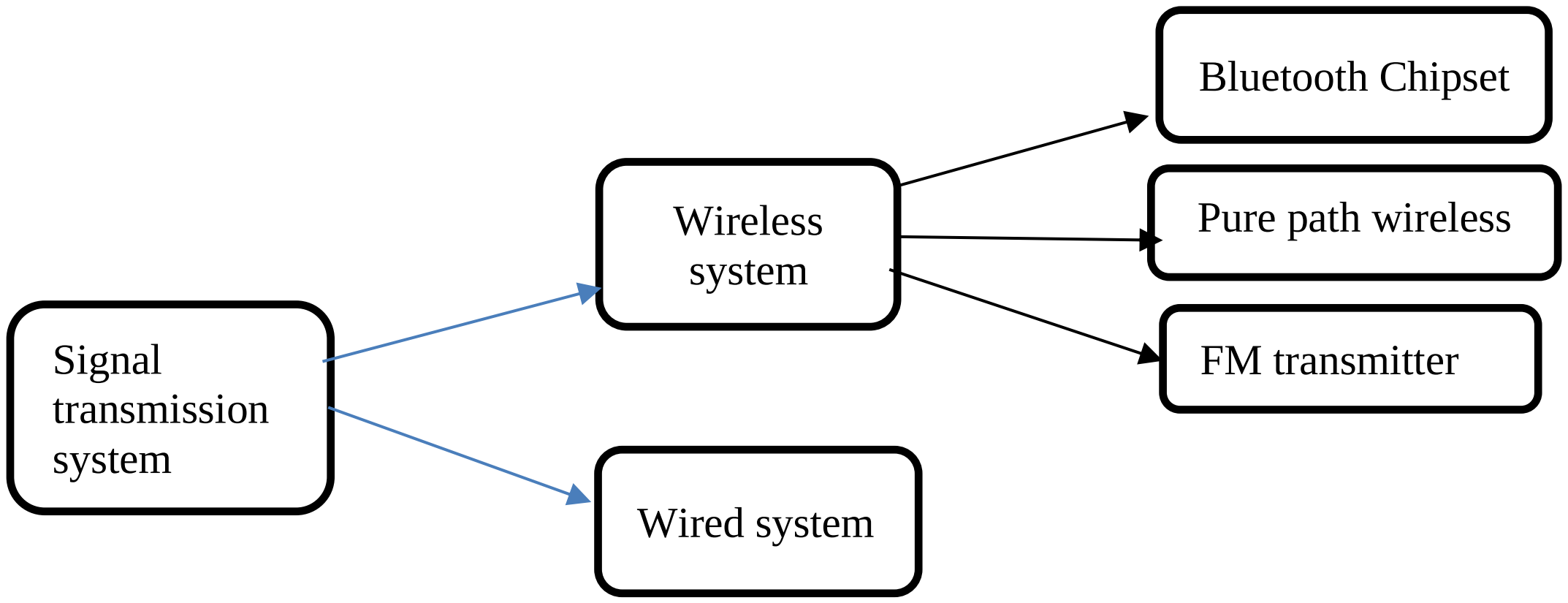

*Fig 7. Classification tree for signal transmission and communication system*

## 4.5.2 Concept Combinations:

The concept combination table provides a way to consider combinations of solution fragments systematically. Here is a combination table that our team used to consider the combinations of fragments of classification trees.

| (Microphones) Transducers | Stethoscope hearing modes | Stethoscope head modes | Stethoscope Processing unit | Sound transmission system | Output Processed data visualization |
|---|---|---|---|---|---|
| Condenser microphone | Electronics | Dual (Diaphragm and bell sides) | None | Tubing(wired) earpieces | None |
| Fiber optics microphone | Acoustic | Single | Filtration and amplification | Bluetooth Chipset | Both LCD and wireless devices |
| MEMS microphone | Dual | | Analog to digital conversion | Pure path Wireless | Embedded LCD Screen |
| None | | | signal recognition and clustering | FM transmitter | |

*Table-1: Concept combination table for digital stethoscope*

The columns in the table correspond to the subproblems identified in concept classification trees. The entries in each column correspond to the solution fragments for each of these subproblems derived from external and internal search. Potential solutions to the overall problem are formed by combining one fragment from each column, there are 1152 possible combinations (4 × 3 × 2 x 4 x 4 x 3). Choosing a combination of fragments does not lead spontaneously to a solution to the overall problem.

**N.B**: All of the group members participated in the concept generation phase. The team members contributed at least two concepts each in the mission category.

- **Concept A: Dual head acoustic Stethoscope**

| Energy Transducers | Stethoscope hearing modes | Stethoscope head modes | Stethoscope Processing unit | Sound transmission system | Output Processed data visualization |
|---|---|---|---|---|---|
| Condenser microphone | Electronics | Single | None | Tubing(wired) earpieces | None |
| Fiber optics microphone | Acoustic | Dual (Diaphragm and bell sides) | Filtration and amplification | Bluetooth Chipset | Both LCD and wireless devices |
| MEMS microphone | Dual mode | | Analog to digital conversion | Pure path Wireless | Embedded LCD Screen |
| None | | | signal recognition and clustering | FM transmitter | |

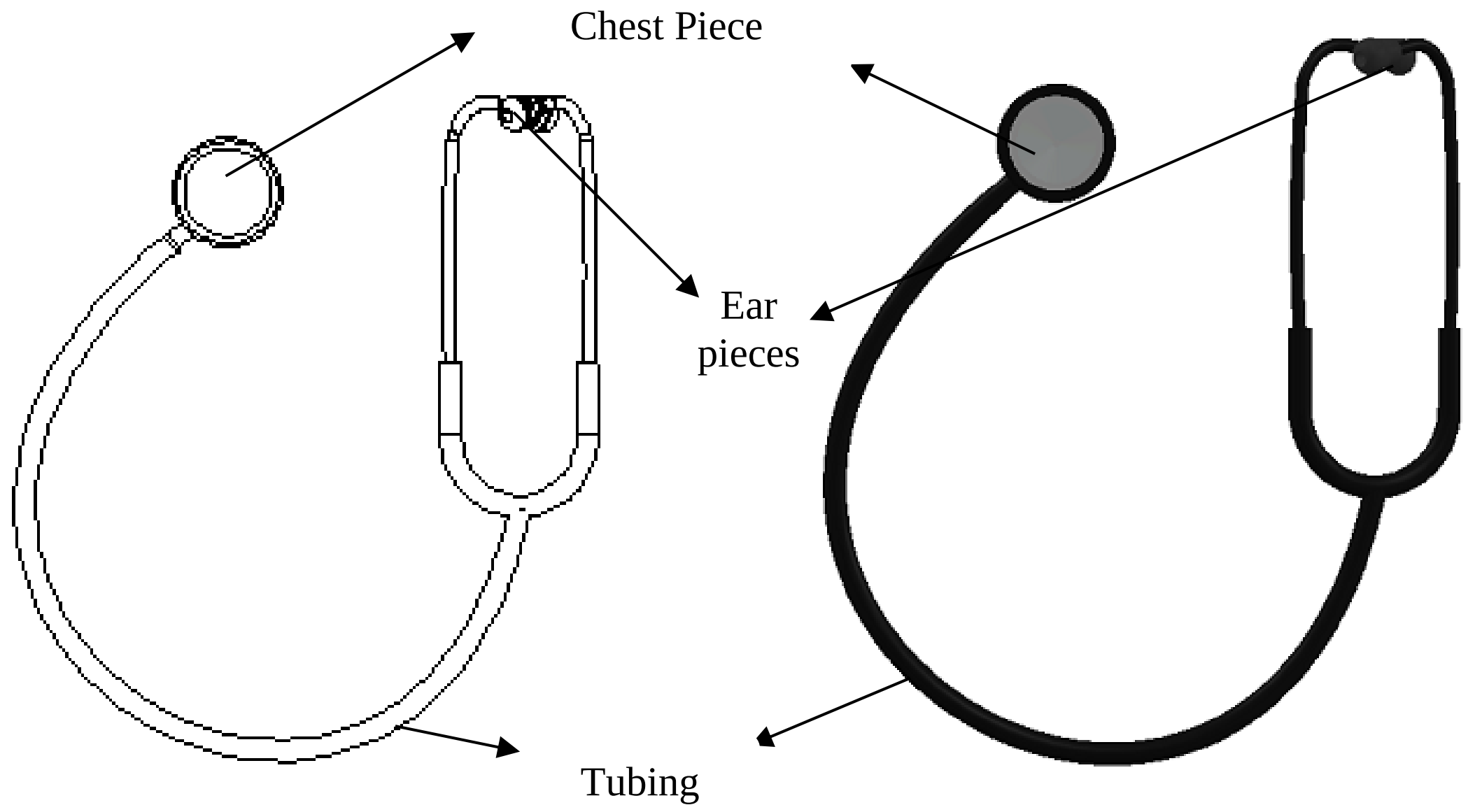

*Fig 8. Concept A product design using SOLIDWORKS software.*

## ❖ Concept B: Analog Electronic Dual Head Stethoscope

| Energy Transducers | Stethoscope hearing modes | Stethoscope head modes | Stethoscope Processing unit | Sound transmission system | Output Processed data visualization |
|---|---|---|---|---|---|
| Condenser microphone | Electronics | Single | None | Tubing(wired) earpieces | None |
| Fiber optics microphone | Acoustic | Dual (Diaphragm and bell sides) | Filtration and amplification | Bluetooth Chipset | Both LCD and wireless devices |
| MEMS microphone | Dual mode | | Analog to digital conversion | Pure path Wireless | Embedded LCD Screen |
| None | | | signal recognition and clustering | FM transmitter | |

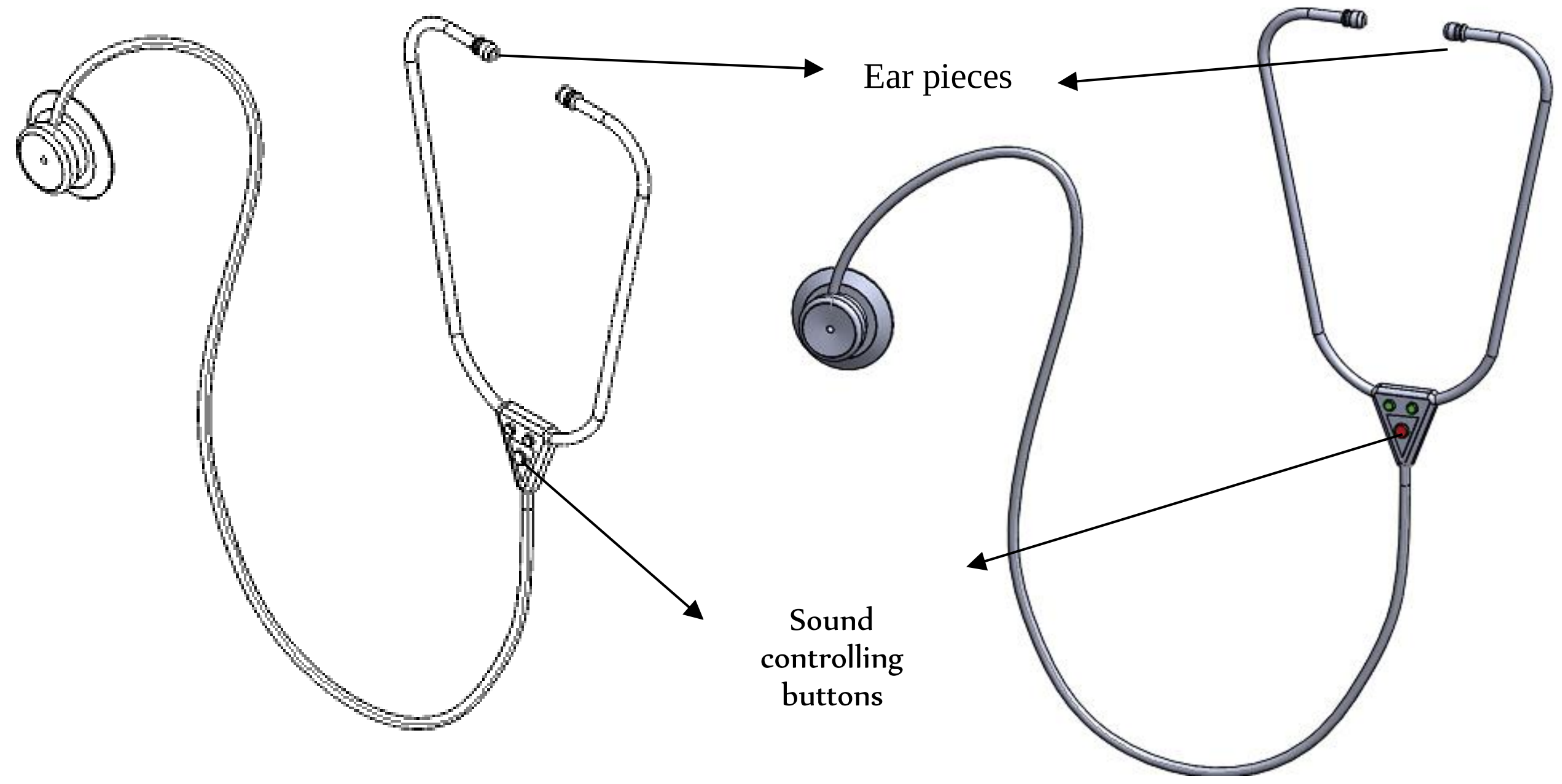

*Fig 9. Concept **B** product design using SOLIDWORKS software.*

## Concept c: Single Head Digital Stethoscope with condenser microphone

| Energy Transducers | Stethoscope hearing modes | Stethoscope head modes | Stethoscope Processing unit | Sound transmission system | Output Processed data visualization |
|---|---|---|---|---|---|
| Condenser microphone | Electronics | Single | None | Tubing(wired) earpieces | None |
| Fiber optics microphone | Acoustic | Dual (Diaphragm and bell sides) | Filtration and amplification | Bluetooth Chipset | Both LCD and wireless devices |
| MEMS microphone | Dual mode | | Analog to digital conversion | Pure path Wireless | Embedded LCD Screen |
| None | | | signal recognition and clustering | FM transmitter | |

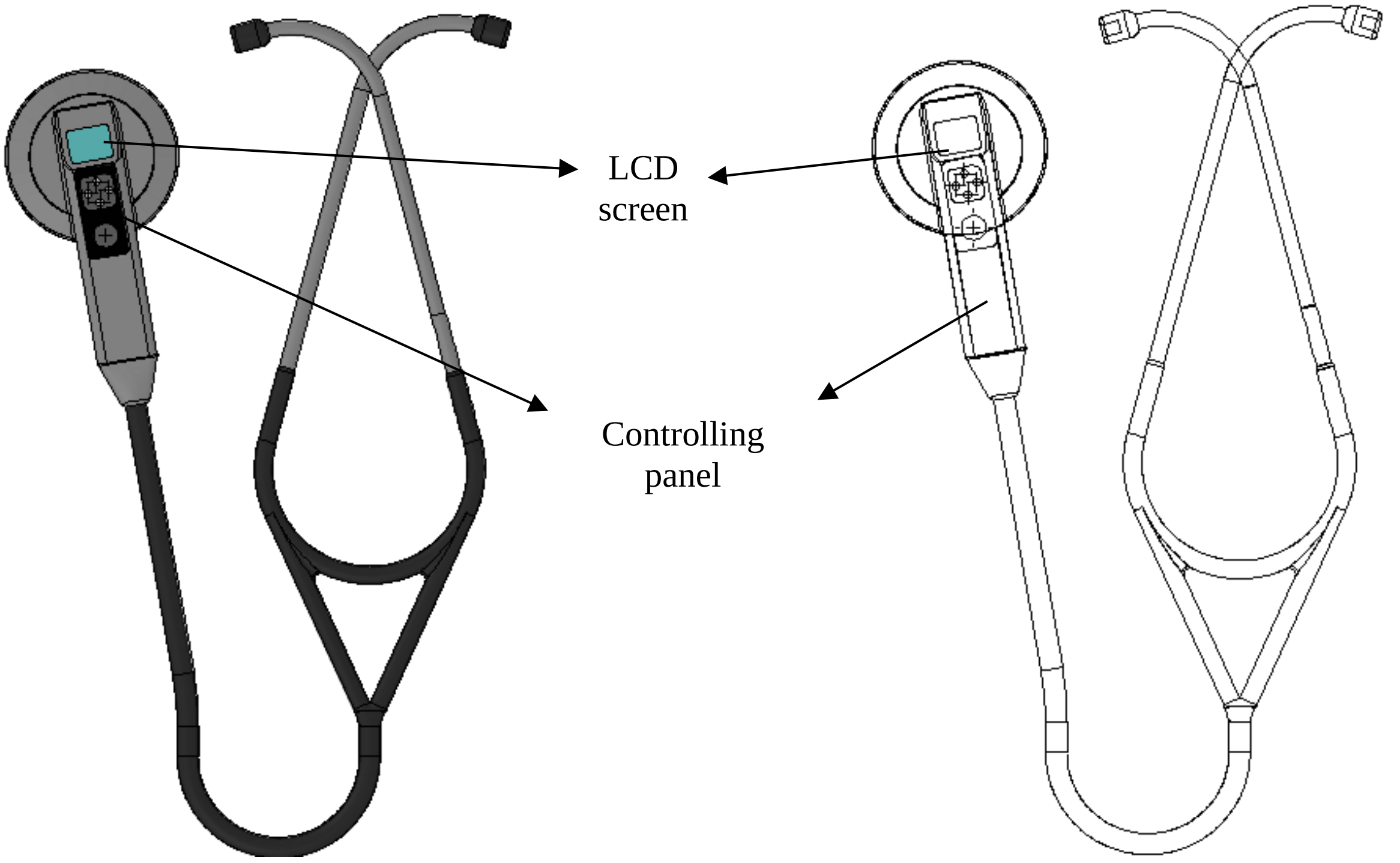

*Fig 10. Concept **C** product design using SOLIDWORKS software.*

## ❖ Concept D: Bluetooth Aided Digital Single Head Stethoscope

| Energy Transducers | Stethoscope hearing modes | Stethoscope head modes | Stethoscope Processing unit | Sound transmission system | Output Processed data visualization |
|---|---|---|---|---|---|
| Condenser microphone | Electronics | Single | None | Tubing(wired) earpieces | None |
| Fiber optics microphone | Acoustic | Dual (Diaphragm and bell sides) | Filtration and amplification | Bluetooth Chipset | Embedded LCD Screen |
| MEMS microphone | Dual mode | | Analog to digital conversion | Pure path Wireless | Both LCD and Bluetooth devices |
| None | | | signal recognition and clustering | FM transmitter | |

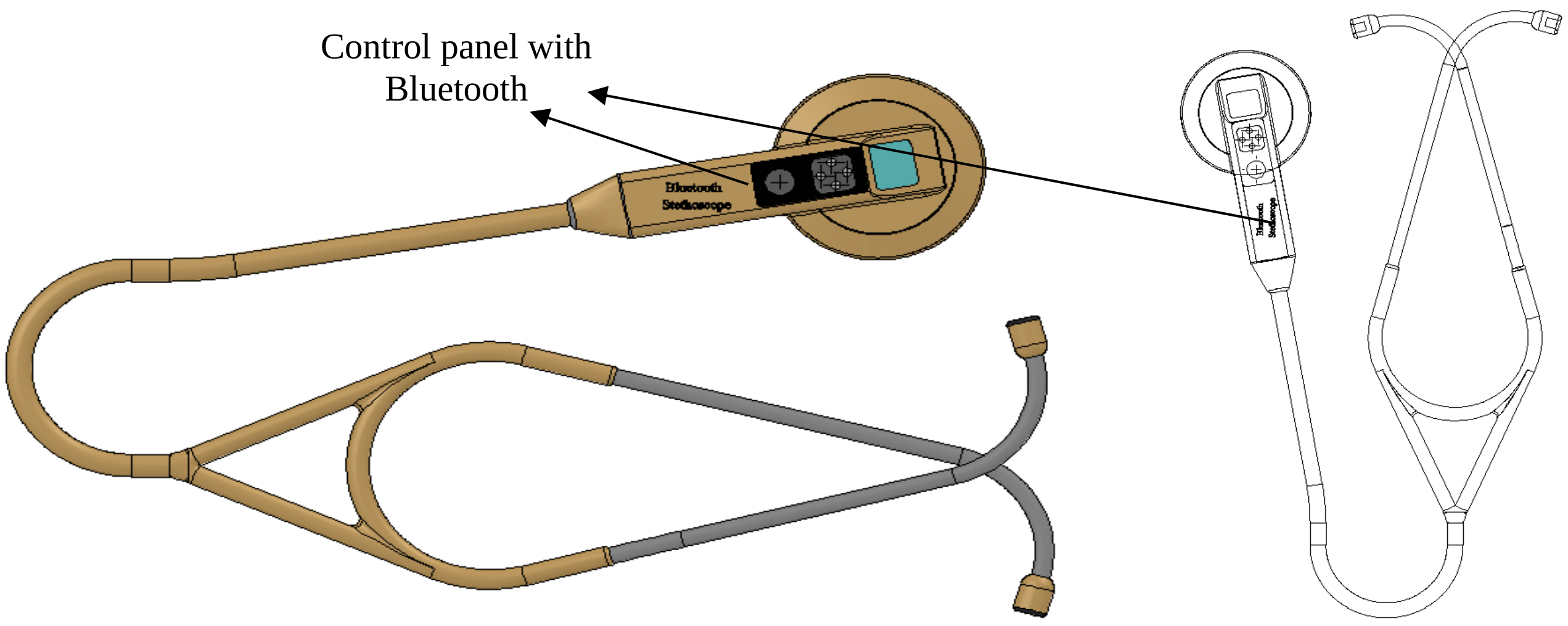

*Fig 11. Concept **D** product design using SOLIDWORKS software.*

❖ **Concept E: Single Head Digital Stethoscope with pure path wireless**

| Energy Transducers | Stethoscope hearing modes | Stethoscope head modes | Stethoscope Processing unit | Sound transmission system | Output Processed data visualization |
|---|---|---|---|---|---|
| Condenser microphone | Electronics | Single | None | Tubing(wired) earpieces | None |
| Fiber optics microphone | Acoustic | Dual (Diaphragm and bell sides) | Filtration and amplification | Bluetooth Chipset | Embedded LCD Screen |
| MEMS microphone | Dual mode | | Analog to digital conversion | Pure path Wireless | Both LCD and Bluetooth devices |
| None | | | signal recognition and clustering | FM transmitter | |

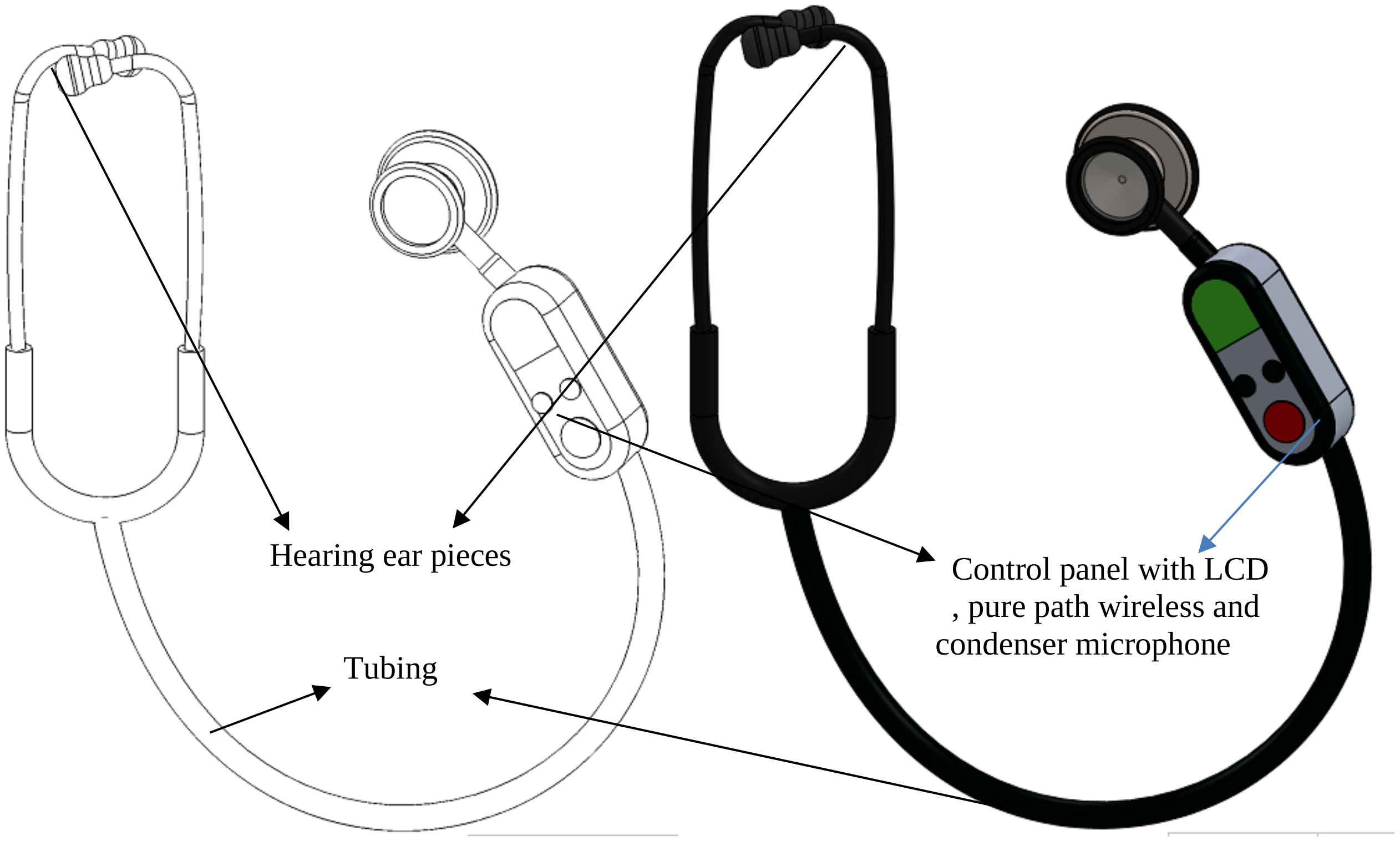

*Fig 12. Concept **E** product design using SOLIDWORKS software.*

❖ **Concept F: AI enabled Digital Stethoscope with FM transmitter.**

| Energy Transducers | Stethoscope hearing modes | Stethoscope head modes | Stethoscope Processing unit | Sound transmission system | Output Processed data visualization |
|---|---|---|---|---|---|
| Condenser microphone | Electronics | Single | None | Tubing(wired) earpieces | None |
| Fiber optics microphone | Acoustic | Dual (Diaphragm and bell sides) | Filtration and amplification | Bluetooth Chipset | Embedded LCD Screen |
| MEMS microphone | Dual mode | | Analog to digital conversion | Pure path Wireless | Both LCD and Bluetooth devices |
| None | | | signal recognition and clustering | FM transmitter | |

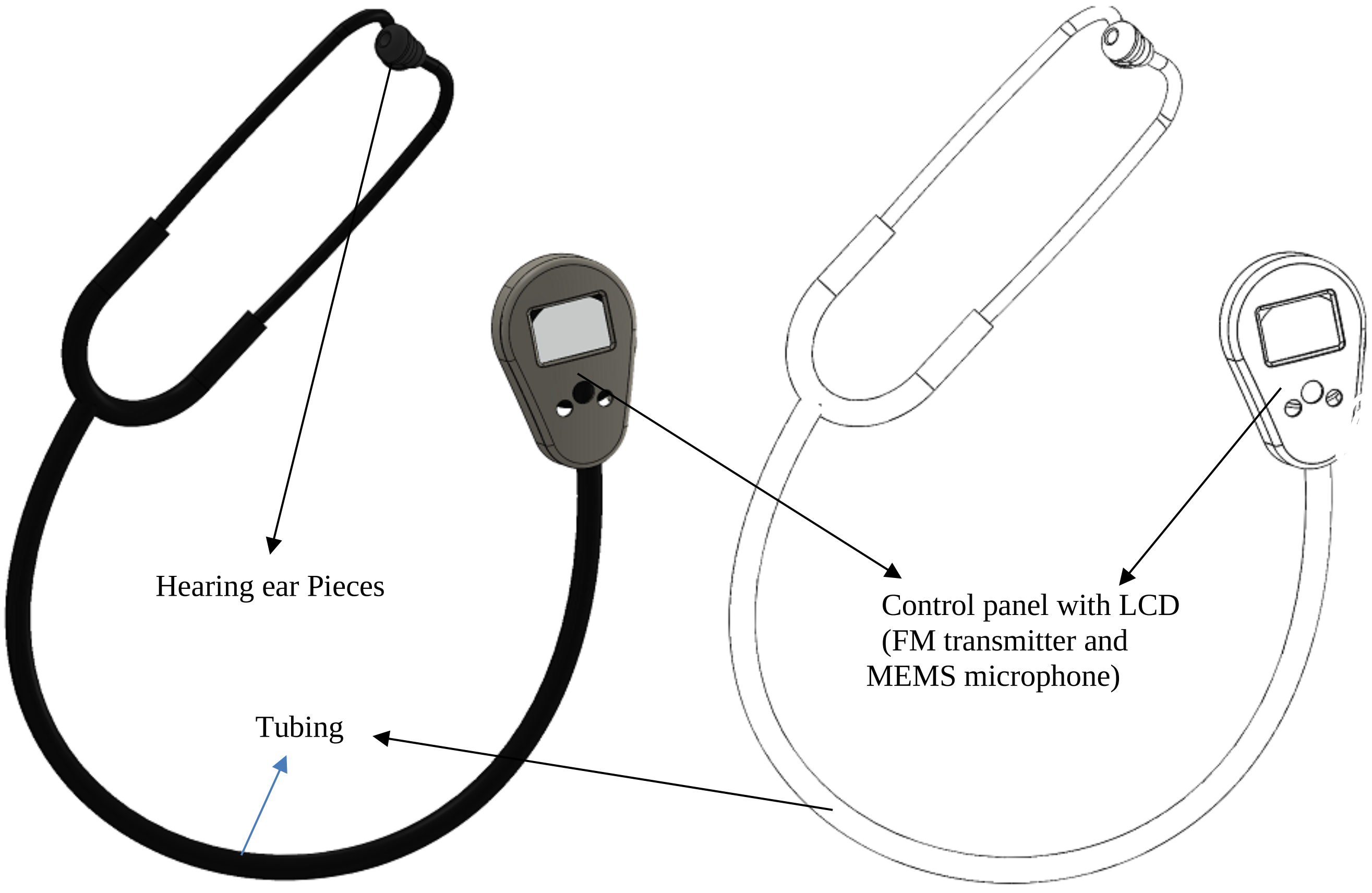

*Fig 13. Concept **F** product design using SOLIDWORKS software.*

All the sketches of the designed stethoscopes under concept generation and conception selection phases are accomplished using the SOLIDWORKS computer program. The sketches are fully worked by the endeavors of the whole group and are claimed to be the property and original version of our team.

## 4.6 Reflect on the Solutions and the Process

In this step there are some questions to check our team done properly the whole process.

- Is the team developing confidence that the solution space has been fully explored? yes
- Are there alternative function diagrams? yes
- Are there alternative ways to decompose the problem? yes
- Have external sources been thoroughly pursued? yes
- Have ideas from everyone been accepted and integrated in the process? yes

# 5) Concept Selection

The process of evaluating concepts with respect to customer needs and other criteria, comparing the relative strengths and weaknesses of the concepts, and selecting one or more concepts for further investigation, testing, or development.

Our team establish a selection criterion for concept selection: those are

- Light weighted (Portability)
- Sensitivity (frequency response)
- Transmission Quality
- Ease of maintenance
- Easy to use
- Output signal quality
- low power consumption
- Cos

From generating concept and customer need we generate the following concepts as follow

- **Concept A: Dual head Acoustic Stethoscope**

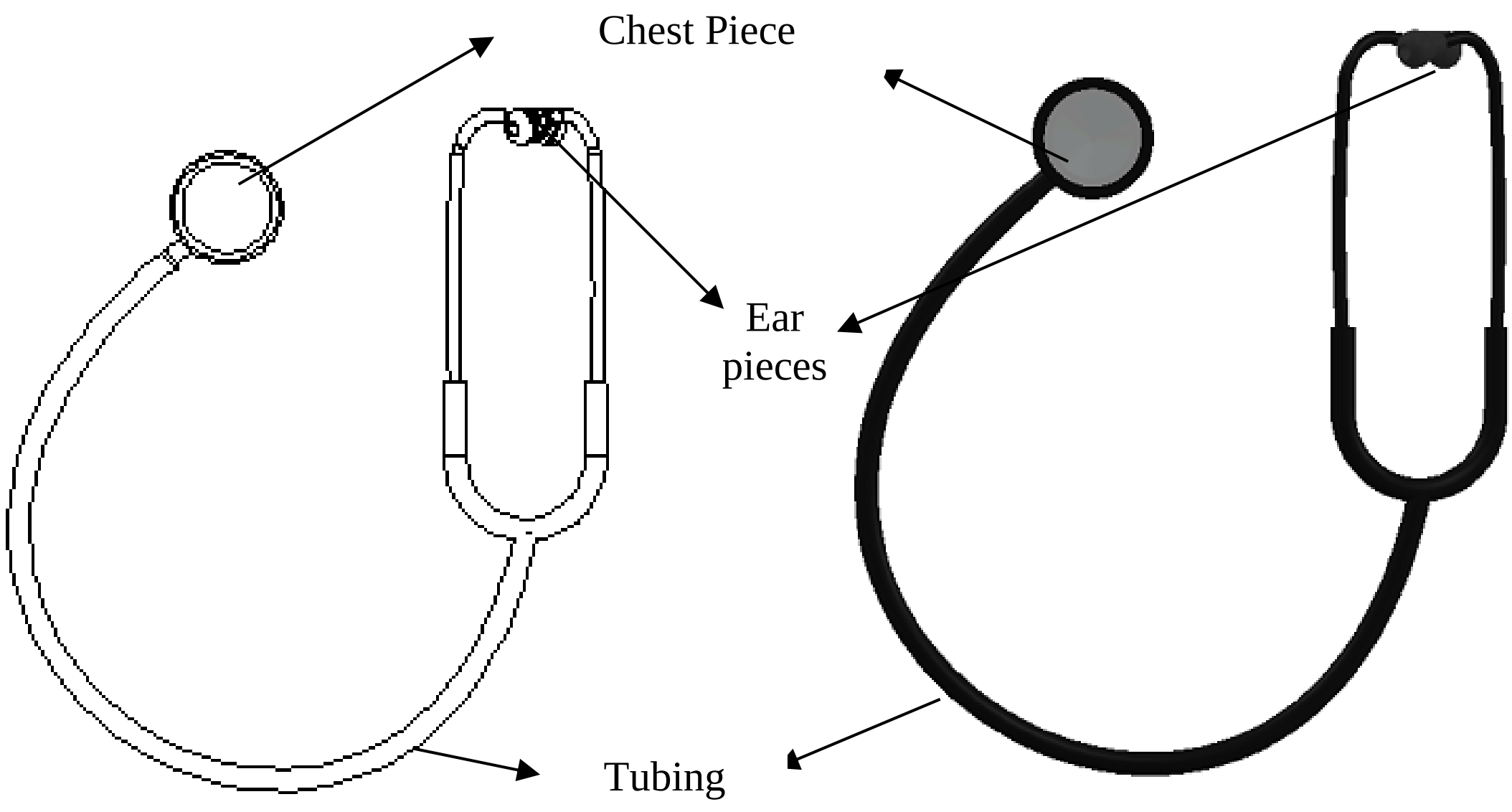

*Fig 1. Concept **A** product design using SOLIDWORKS software.*

- **Concept B: Analog Electronic Dual Head Stethoscope**

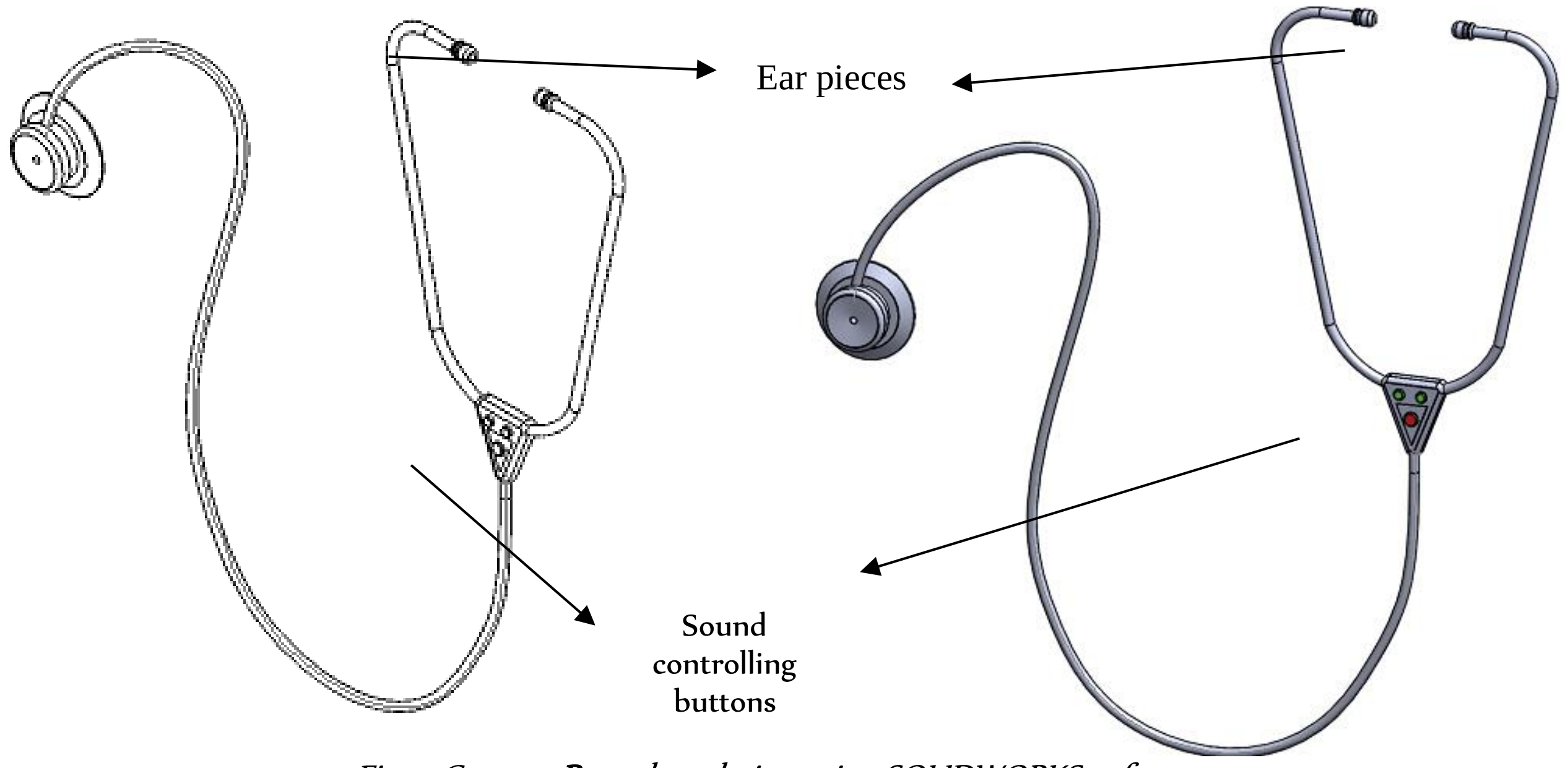

*Fig 2. Concept **B** product design using SOLIDWORKS software.*

- **Concept C: Single Head Digital Stethoscope with condenser microphone**

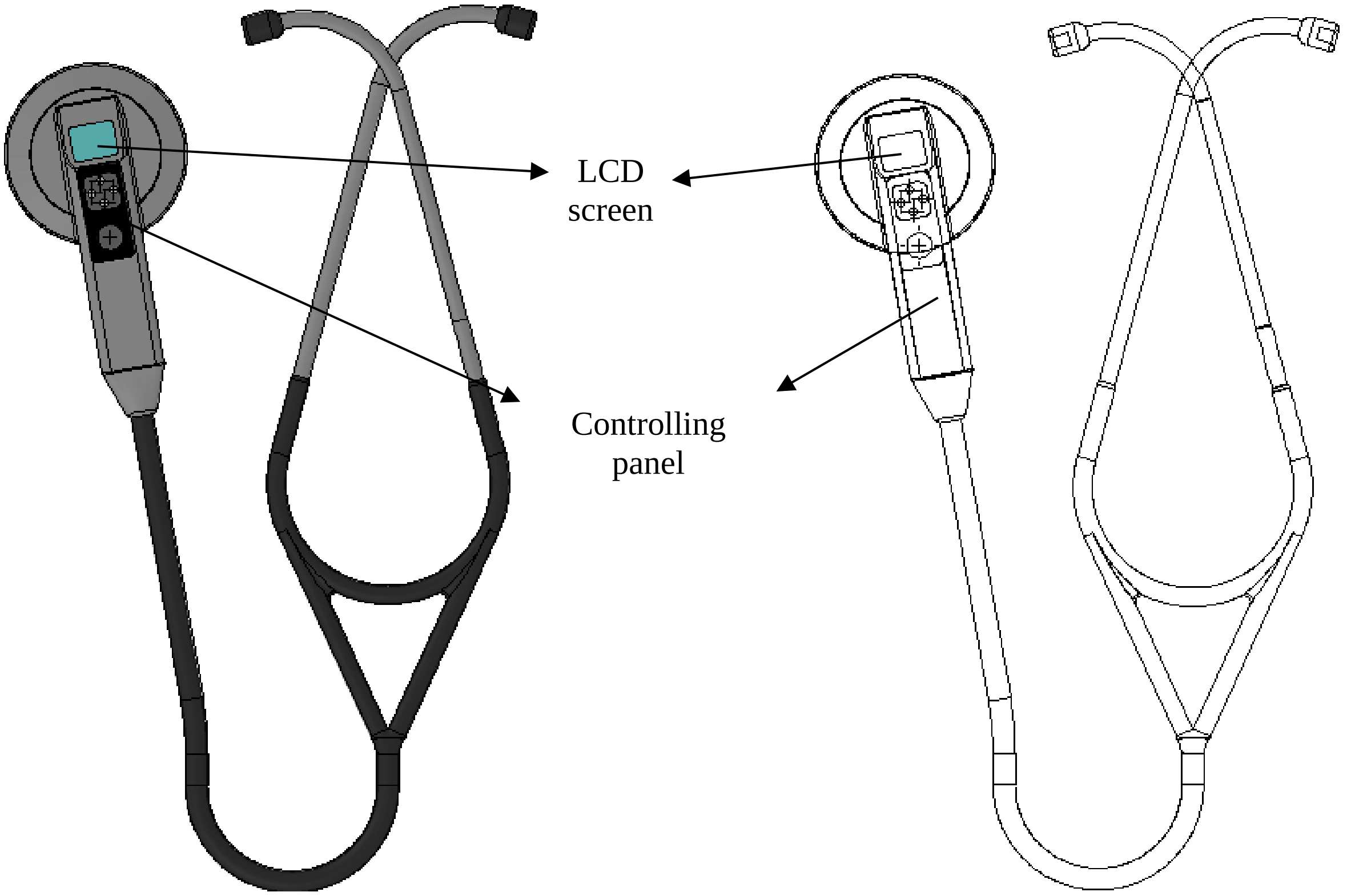

*Fig 3. Concept **C** product design using SOLIDWORKS software.*

- **Concept D: Bluetooth Aided Digital Single Head Stethoscope**

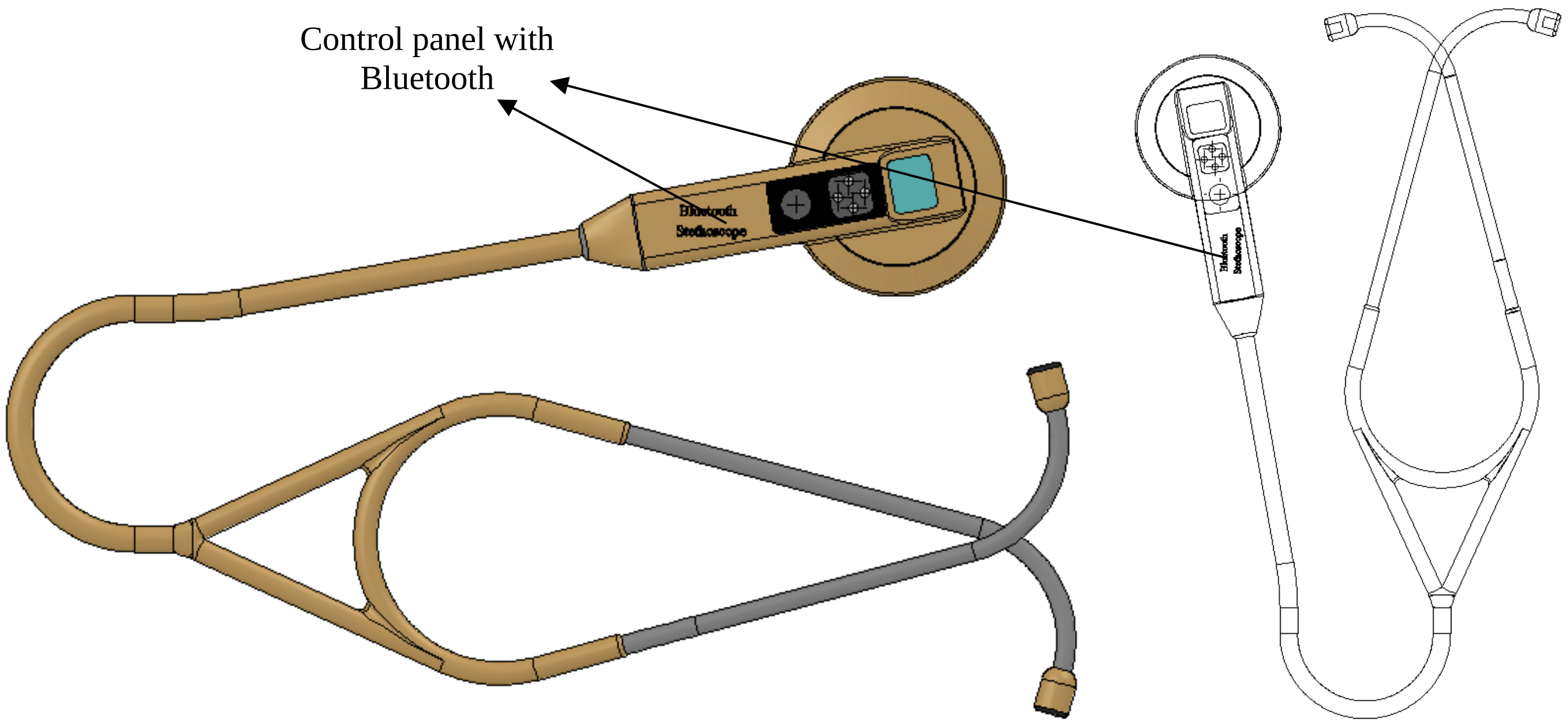

*Fig 4. Concept **B** product design using SOLIDWORKS software.*

- **Concept E: Single Head Digital Stethoscope with pure path wireless**

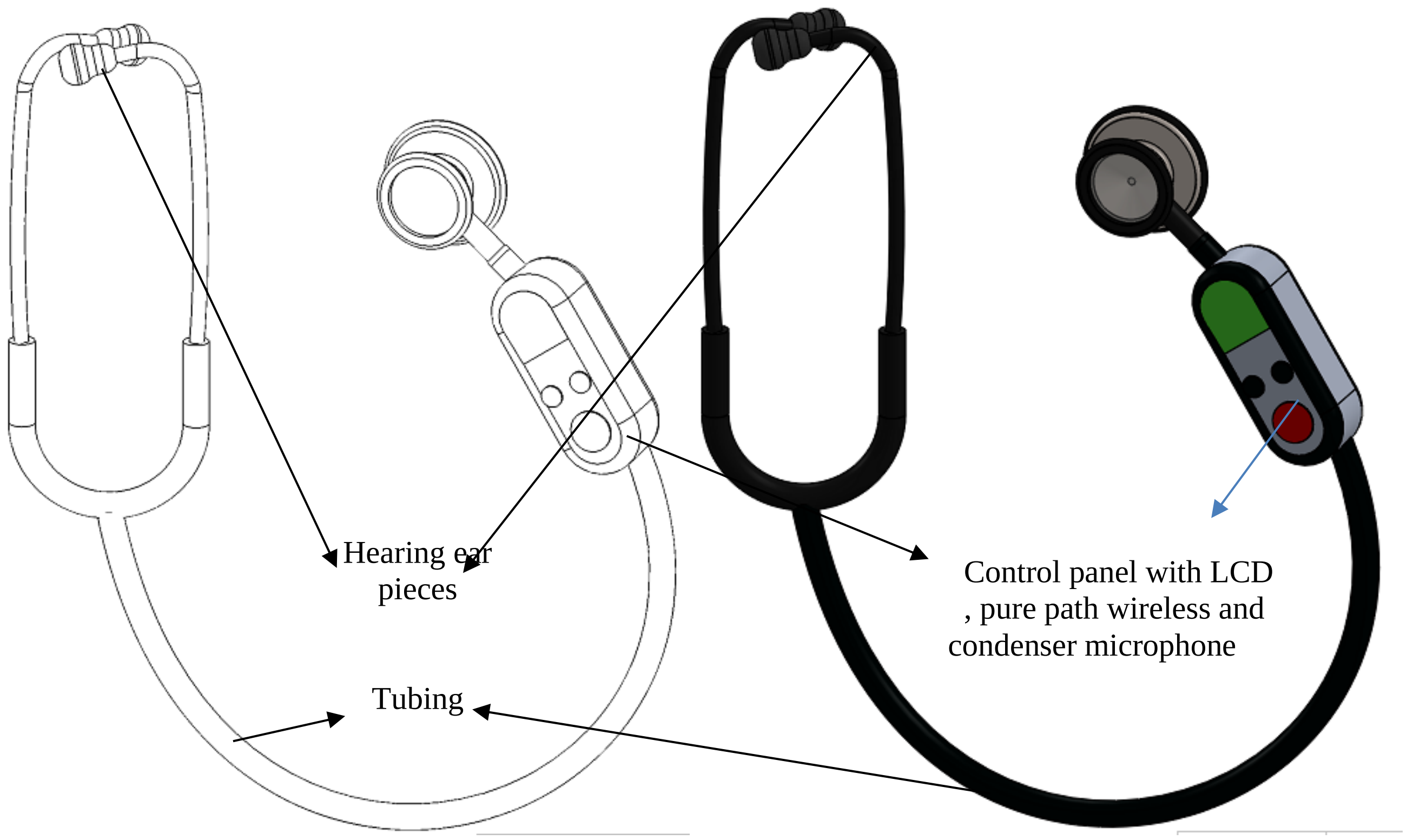

*Fig 5. Concept **E** product design using SOLIDWORKS software.*

- **Concept F: Digital Stethoscope with AI embedded with FM transmitter.**

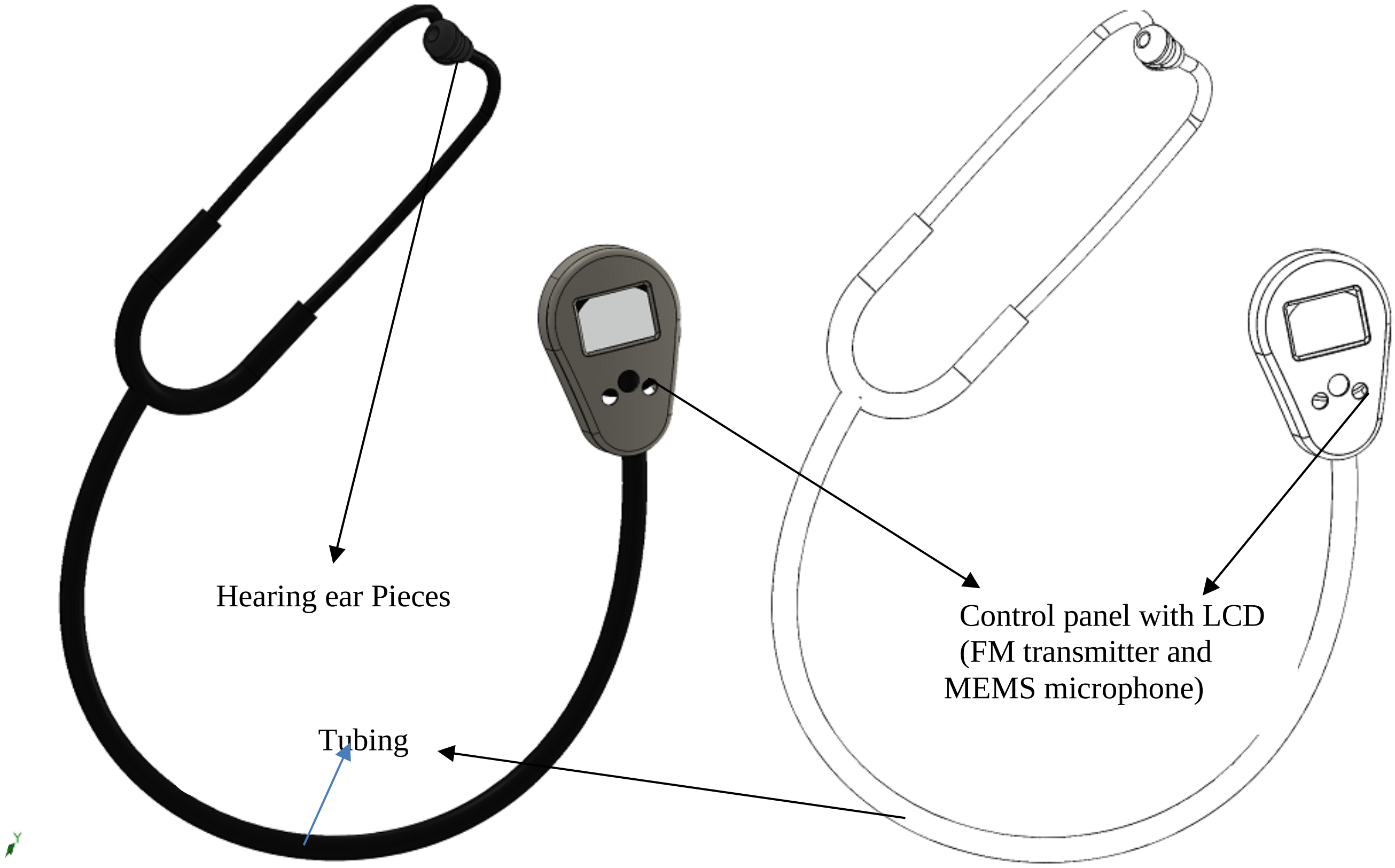

*Fig 6. Concept **B** product design using SOLIDWORKS software.*

## 5.1 Concept Screening

The purposes of this stage are to narrow the number of concepts quickly and to improve the concepts.

### 5.1.1 prepare selection matrix

To prepare the matrix, the team selects a physical medium appropriate to the problem at hand. Individuals and small groups with a short list of criteria may use matrices below.

| | Concepts | | | | | |
|---|---|---|---|---|---|---|
| Selection criteria | Concept A | Concept B (Reference) | Concept C | Concept D | Concept E | Concept F |
| Light weighted (Portability) | | | | | | |
| Sensitivity | | | | | | |
| Transmission Quality | | | | | | |
| Ease of maintenance | | | | | | |
| Easy to use | | | | | | |
| Signal quality(output) | | | | | | |
| low power consumption | | | | | | |
| Cost | | | | | | |
| Sum +'s | | | | | | |
| Sum 0's | | | | | | |
| Sum -'s | | | | | | |
| Net score | | | | | | |
| rank | | | | | | |
| Continue? | | | | | | |

*Table-1: Concept screening selection matrix table for digital stethoscope*

### 5.1.2 Rate selection matrix

A relative score of "better than" (+), "same as" (0), or "worse than" (–) is placed in each cell of the matrix to represent how each concept rates in comparison to the reference concept relative to the particular criterion.

| | Concepts | | | | | |
|---|---|---|---|---|---|---|
| Selection criteria | Concept A | Concept B (Reference) | Concept C | Concept D | Concept E | Concept F |
| Light weighted (Portability) | 0 | 0 | - | - | - | 0 |
| Sensitivity | - | 0 | - | + | + | + |
| Transmission Quality | - | 0 | 0 | + | 0 | 0 |
| Ease of maintenance | 0 | 0 | 0 | + | + | 0 |
| Easy to use | - | 0 | + | + | + | + |
| Signal quality(output) | - | 0 | 0 | 0 | 0 | + |
| low power consumption | + | 0 | 0 | 0 | 0 | 0 |
| Cost | 0 | 0 | 0 | - | - | - |
| Sum +'s | | | | | | |
| Sum 0's | | | | | | |
| Sum -'s | | | | | | |
| Net score | | | | | | |
| rank | | | | | | |
| Continue? | | | | | | |

*Table-2: Concept screening selection rate table for digital stethoscope*

## 5.1.3 Rank selection matrix

After rating all the concepts, the team sums the number of "better than," "same as," and "Worse than" scores and enters the sum for each category in the lower rows of the matrix. Once the summation is completed, the team rank-orders the concepts. Obviously, in general those concepts with more pluses and fewer minuses are ranked higher

| | Concepts | | | | | |
|---|---|---|---|---|---|---|
| Selection criteria | Concept A | Concept B (Reference) | Concept C | Concept D | Concept E | Concept F |
| Light weighted (Portability) | 0 | 0 | - | - | - | 0 |
| Sensitivity (frequency response) | - | 0 | - | + | + | + |
| Transmission Quality | - | 0 | 0 | + | 0 | 0 |
| Ease of maintenance | 0 | 0 | 0 | + | + | 0 |
| Easy to use | - | 0 | + | + | + | + |
| Signal quality(output) | - | 0 | 0 | 0 | 0 | + |
| low power consumption | + | 0 | 0 | 0 | 0 | 0 |
| Cost | 0 | 0 | 0 | - | - | - |
| Sum +'s | 1 | 0 | 1 | 4 | 3 | 3 |
| Sum 0's | 3 | 8 | 5 | 2 | 3 | 4 |
| Sum -'s | 4 | 0 | 2 | 2 | 2 | 1 |
| Net score | -3 | 0 | -1 | 2 | 1 | 2 |
| rank | 6 | 4 | 5 | 1 | 3 | 1 |
| Continue? | no | no | no | yes | yes | yes |

*Table-3: Concept screening rank selection table for digital stethoscope*

## 5.1.4 combine and improve the concepts

Our team verify that the result make sense and then consider some concepts are combined but we noticed that combined concepts couldn't remove several of the "worse than" ratings to yield a new concept,

Two issues are considered:

I. Is there a generally good concept that is degraded by one bad feature? Can a minor modification improve the overall concept and yet preserve a distinction from the other concepts?

II. Are there two concepts that can be combined to preserve the "better than" qualities while annulling the "worse than" qualities?

Combined and improved concepts are then added to the matrix, rated by the team, and ranked along with the original concepts. Our team noticed that concepts D and F could be combined to remove several of the "worse than" ratings to yield a new concept, DF, to be considered in the next round.

❖ **Concept DF: Advanced Digital Stethoscope with AI embedded.**

| Energy Transducers | Stethoscope hearing modes | Stethoscope head modes | Stethoscope Processing unit | Sound transmission system | Output Processed data visualization |
|---|---|---|---|---|---|
| Condenser microphone | Electronics | Single | None | Tubing(wired) earpieces | Wireless Devices |
| Fiber optics microphone | Analog | Dual (Diaphragm and bell sides) | Filtration and amplification | Bluetooth Chipset | Embedded LCD Screen |
| MEMS microphone | Dual mode | | Analog to digital conversion | Pure path Wireless | Both LCD and Bluetooth devices |
| None | | | signal recognition and clustering | FM transmitter | None |

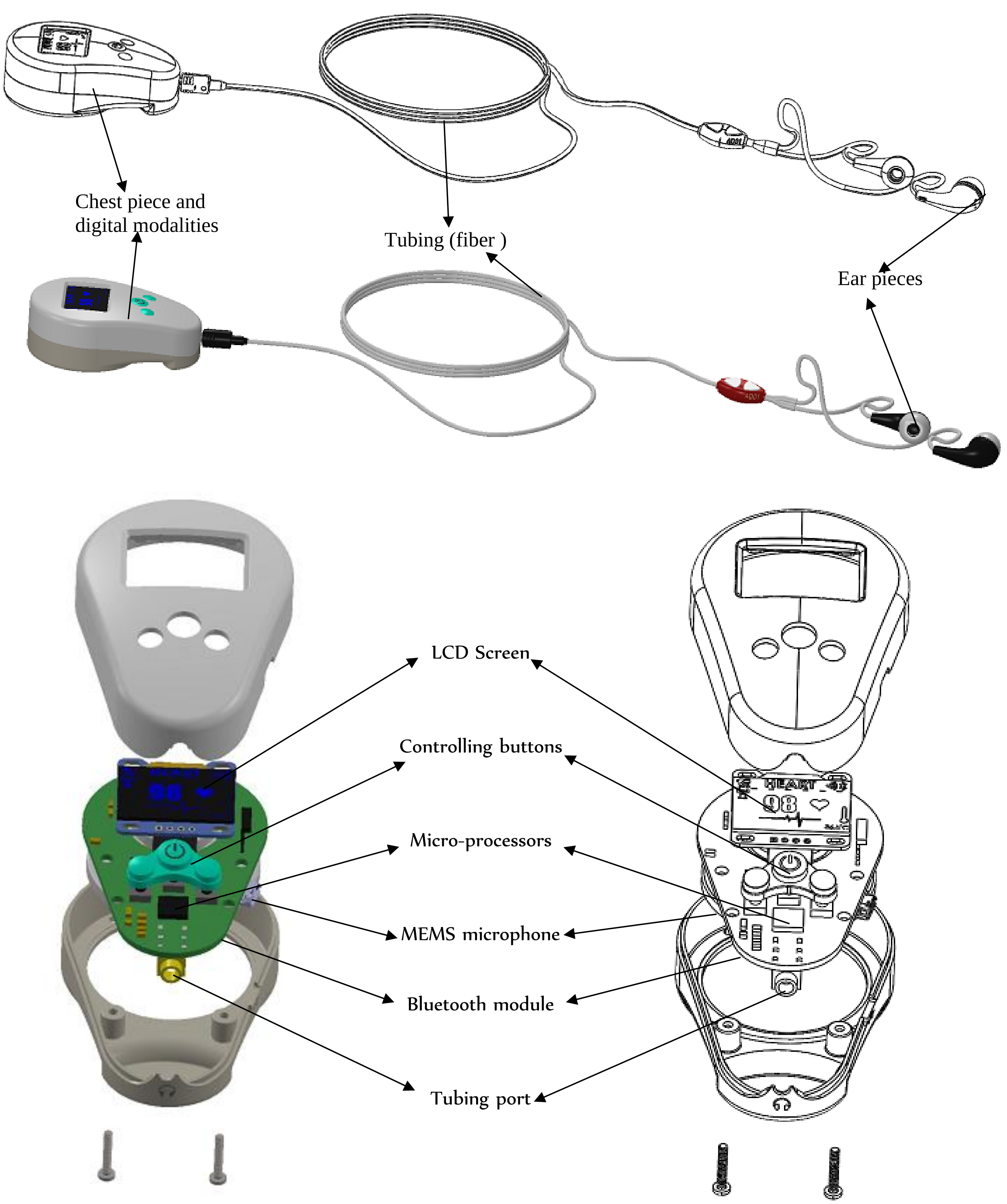

*Fig 7: Combined concept product design using SOLIDWORKS software*

### 5.1.5 select one or more concepts

From concept screening our team decide concept D, E, F and DF are to be select for further refinement and analysis.

### 5.1.6 reflect on the results and the process

Our team members are comfortable with the outcome and agrees on the whole process of concept screening

## 5.2 Concept Scoring

Concept scoring is used when increased resolution will better differentiate among competing concepts. In this stage, the team weighs the relative importance of the selection criteria and focuses on refined comparisons with respect to each criterion. The concept scores are determined by the weighted sum of the ratings.

- **Step 1: Prepare the Selection Matrix**

The team prepares a matrix and identifies a reference concept. In most cases a computer spreadsheet is the best format to facilitate ranking and sensitivity analysis.

| | | Concept | | | | | | | |
|---|---|---|---|---|---|---|---|---|---|
| | | D | | E | | F | | DF | |
| Selection criteria | weight | rating | Weight score | rating | Weight score | rating | Weight score | rating | weight Score |
| Light weighted (Portability) | | | | | | | | | |
| Sensitivity (frequency response) | | | | | | | | | |
| Transmission Quality | | | | | | | | | |
| Ease of maintenance | | | | | | | | | |
| Easy to use | | | | | | | | | |
| Signal quality(output) | | | | | | | | | |
| low power consumption | | | | | | | | | |
| Low Cost | | | | | | | | | |
| | Total/rank | | | | | | | | |
| | continue | | | | | | | | |

*Table-4: Concept scoring selection matrix table*

- **Step 2: Rate the Concepts**

We are rating the concept first by rating the selection criteria using finer scale. This is as follow

| Relative performance | rating |
|---|---|
| Much worse than reference | 1 |
| Worse than reference | 2 |
| Same as reference | 3 |
| Better than reference | 4 |
| Much better than reference | 5 |

| | | Concept | | | | | | | |
|---|---|---|---|---|---|---|---|---|---|
| | | D | | E | | F | | DF | |
| Selection criteria | weight | rating | Weight | rating | Weight | rating | Weight | rating | weight |
| | | | score | | score | | score | | Score |
| Light weighted (Portability) | 0.1 | 3 | 0.3 | 3 | 0.3 | 4 | 0.4 | 4 | 0.4 |
| Sensitivity (frequency response) | 0.15 | 4 | 0.6 | 4 | 0.6 | 4 | 0.6 | 4 | 0.6 |
| Transmission Quality | 0.15 | 4 | 0.6 | 5 | 0.75 | 3 | 0.45 | 4 | 0.6 |
| Ease of maintenance | 0.1 | 5 | 0.5 | 4 | 0.4 | 4 | 0.4 | 5 | 0.5 |
| Easy to use | 0.15 | 4 | 0.6 | 4 | 0.6 | 5 | 0.75 | 5 | 0.75 |
| Signal quality(output) | 0.2 | 3 | 0.6 | 3 | 0.6 | 5 | 1.00 | 5 | 1.00 |
| low power consumption | 0.05 | 3 | 0.15 | 2 | 0.1 | 4 | 0.2 | 4 | 0.2 |
| Cost | 0.1 | 4 | 0.4 | 4 | 0.4 | 3 | 0.3 | 3 | 0.3 |
| Total Score | | | | | | | | | |
| rank | | | | | | | | | |
| continue | | | | | | | | | |

*Table-5: Concept Scoring selection rate table*

The team decided that the criterion "ease of use" is considered to provide general detail of usage. To clarify it more "Ease of use "(*weight= 0.15*) could be broken down, as shown in, to include "ease of loading," "ease of reading and reading," and "ease of cleaning", and each weighting *0.05.*

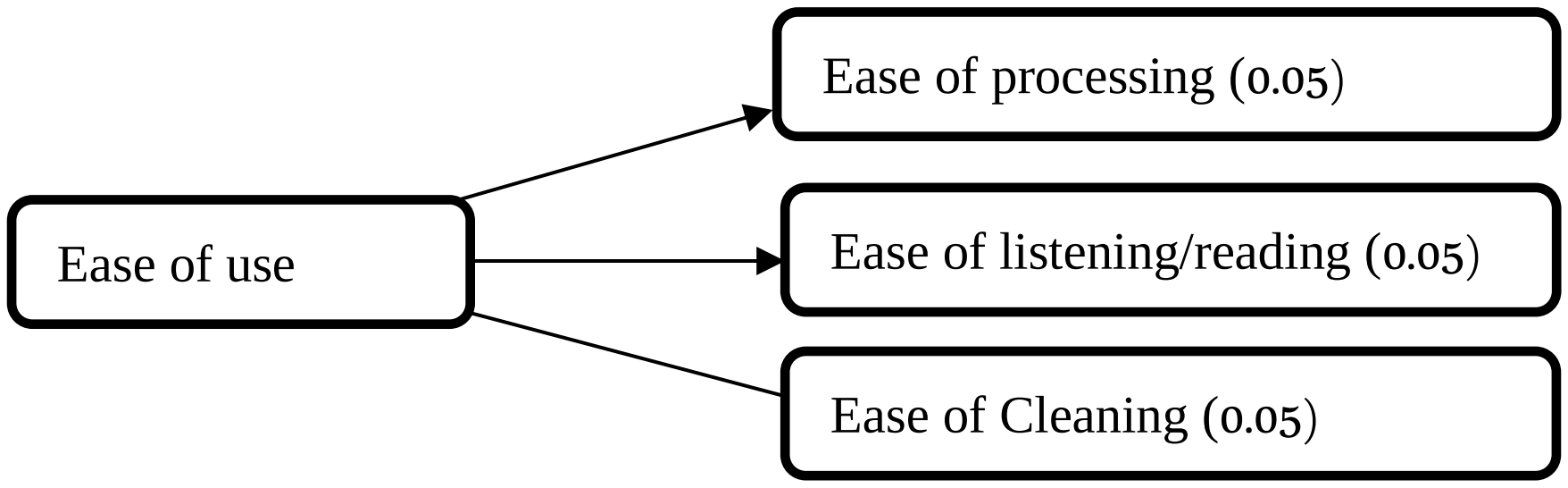

- **Step 3: Rank the Concepts**

Once the ratings are entered for each concept, weighted scores are calculated by multiplying the raw scores by the criteria weights. total score for each concept is the sum of the weighted scores.

| | | Concept | | | | | | | |
|---|---|---|---|---|---|---|---|---|---|
| | | D | | E | | F | | DF | |
| Selection criteria | weight | rating | Weight score | rating | Weight score | rating | Weight score | rating | weight Score |
| Light weighted (Portability) | 0.1 | 3 | 0.3 | 3 | 0.3 | 4 | 0.4 | 4 | 0.4 |
| Sensitivity (frequency response) | 0.15 | 4 | 0.6 | 4 | 0.6 | 4 | 0.6 | 4 | 0.6 |
| Transmission Quality | 0.15 | 4 | 0.6 | 5 | 0.75 | 3 | 0.45 | 4 | 0.6 |
| Ease of maintenance | 0.1 | 5 | 0.5 | 4 | 0.4 | 4 | 0.4 | 5 | 0.5 |
| Easy to use | 0.15 | 4 | 0.6 | 4 | 0.6 | 5 | 0.75 | 5 | 0.75 |
| Signal quality(output) | 0.2 | 3 | 0.6 | 3 | 0.6 | 5 | 1.00 | 5 | 1.00 |
| low power consumption | 0.05 | 3 | 0.15 | 2 | 0.1 | 4 | 0.2 | 4 | 0.2 |
| Cost | 0.1 | 4 | 0.4 | 4 | 0.4 | 3 | 0.3 | 3 | 0.3 |
| Total Score | | 3.75 | | 3.45 | | 4.1 | | 4.35 | |
| rank | | 3 | | 4 | | 2 | | 1 | |
| continue | | No | | No | | No | | Develop | |

*Table-6: Concept ranking table of Concept scoring process*

- **Step 4: Combine and Improve the Concepts**

Our team noticed that combined concepts couldn't remove several of the "worse than" ratings to yield a new concept.

- **Step 5: Select One or More Concepts**

The team considered carefully the significance of differences in concept scores. Given the resolution of the scoring system, small differences are generally not significant. From concept scoring our team agreed that *concept DF* is the most promising and would be likely to result in a successful product.

- **Step 6: Reflect on the Results and the Process**

Our team members are following the whole process of concept scoring and concept selection process and agree to each final result of each process and Our team members feels comfortable that all of the relevant issues have been discussed and that the selected concept(s) have the greatest potential to satisfy customers and be economically successful.